\documentclass[twoside,twocolumn,english,pra,show pacs]{revtex4}
\usepackage[T1]{fontenc}
\usepackage[latin9]{inputenc}
\usepackage{color}
\usepackage{amsmath}
\usepackage{amssymb}
\usepackage{graphicx}
\usepackage{esint}

\makeatletter
\@ifundefined{textcolor}{}
{%
 \definecolor{BLACK}{gray}{0}
 \definecolor{WHITE}{gray}{1}
 \definecolor{RED}{rgb}{1,0,0}
 \definecolor{GREEN}{rgb}{0,1,0}
 \definecolor{BLUE}{rgb}{0,0,1}
 \definecolor{CYAN}{cmyk}{1,0,0,0}
 \definecolor{MAGENTA}{cmyk}{0,1,0,0}
 \definecolor{YELLOW}{cmyk}{0,0,1,0}
 }

\usepackage{multirow}

\makeatother

\usepackage{babel}
\begin{document}

\title{Conductivity of strongly correlated bosons in optical lattices \\in an Abelian synthetic magnetic field}

\author{A. S. Sajna, T. P. Polak, and R. Micnas}
\begin{abstract}
Topological phase engineering of neutral bosons loaded in an optical
lattice opens a new window for manipulating of transport phenomena
in such systems. Exploiting the Bose Hubbard model and using the magnetic
Kubo formula proposed in this paper we show that the optical conductivity
abruptly changes for different flux densities in the Mott phase. Especially,
when the frequency of the applied field corresponds to the on-site
boson interaction energy, we observe insulator or metallic behavior
for a given Hofstadter spectrum. We also prove, that for different
synthetic magnetic field configurations, the critical conductivity
at the tip of the lobe is non-universal and depends on the energy
minima of the spectrum. In the case of $1/2$ and $1/3$ flux per
plaquette, our results are in good agreement with those of the previous
Monte Carlo (MC) study. Moreover, we show that for half magnetic-flux
through the cell the critical conductivity suddenly changes in the
presence of a superlattice potential with uniaxial periodicity.
\end{abstract}


\address{Solid State Theory Division, Faculty of Physics, A. Mickiewicz University,
ul. Umultowska 85, 61-614 Pozna\'{n}, Poland}

\maketitle

\section{Introduction}

The Bose Hubbard model (BHM) is commonly used to describe many interesting
physical systems e.g. superconductors with short coherence length
\cite{Micnas1990}, Josephson junction arrays \cite{Cha1994,VanOtterloA1993}
or quantum phase transitions in cold quantum gases \cite{Greiner2002,Stoferle2004}
on which the idea of quantum simulations can be realized \cite{Jaksch2005}.
Recently, attention has been paid to the BHM behavior in a strong
synthetic magnetic field \cite{Saha2010,Powell2010,Zaleski2011,Powell2011,Sinha2011,Nakano2012,Polak2013,Struck2013}
as well as to its transport properties \cite{Smakov2005,Bhaseen2007,Bhaseen2009,Lindner2010,Tokuno2011,Chen2013,Kessler}.
This research has opened the possibility of highly controllable BHM
dynamics with explicitly designed kinetics which is a subject of our
study.

Despite of the Josephson junction arrays \cite{Zant1992,VanderZantHS1996}
up to now a strong magnetic field regime has been available by simulating
a vector potential imposed on many body wave function of ultra-cold
neutral gases. It has been realized by engineering the adequate phase
of atoms when they change their quantum states upon hopping through
the lattice sites, using e.g. Raman- and photo-assisted tunneling
\cite{Lin2009a,Aidelsburger2011,Aidelsburger2013,Miyake2013a}, shaking
of the lattice \cite{Struck2011,Struck2013}. In particular, staggered
\cite{Aidelsburger2011} and uniform \cite{Aidelsburger2013,Miyake2013a}
magnetic field have been created. Such synthetic magnetic field has
also been proposed to be generated by combining quadrupolar potential
and modulation of tunneling in time \cite{Sorensen2005c}, although
it has not been realized yet (see also \cite{Jaksch2003}). Moreover,
the relevant gauge degrees of freedom can be precisely experimentally
verifiable and can cause interesting effects like finite momentum
condensate \cite{Polak2013}. As follows from the above, the possibility
of simulating a vector potential provides many opportunities which
has made it a rapidly expanding area of research. The interest in
the simulation has been growing in view of possible future application
of ultra-cold quantum gases in topological quantum computation or
spintronics \cite{Nayak2008a,Hasan2010}.

If we consider the BHM equipped with orbital effects, its complexity
considerably increases, as it develops a multi-subband structure dependent
on the quantity of flux per plaquette. To the best of our knowledge
the conductivity in a strong magnetic field has been considered mainly
numerically in just a few papers. One of the main problems in carrying
out the related calculations is the complex hopping term (Peierls
factor) which occurs in the BHM Hamiltonian. This is the reason why
the optical conductivity (OC) has been rarely studied. Y. Nishiyama
\cite{Nishiyama2001} has analyzed OC in the hard-core limit of the
BHM with disorder. OC in Josephson junction arrays has been also studied
within Landau levels framework without commensurability effects of
magnetic field \cite{VanOtterloA1993,Wagenblast1997}. The critical
conductivity could be much easier available in numerical calculations
thanks to the correspondence between the BHM at integer filling and
the XY model \cite{Cha1994,Kim2008} in which the cases $f=1/2$ and
$f=1/3$ flux per plaquette has been studied \cite{Cha1994}. Using
this correspondence E. Granato and J. Kosterlitz \cite{Granato1990}
derived analytically the value of critical conductivity for half flux
per plaquette. Also Y. Nishiyama \cite{Nishiyama2001} by applying
the exact diagonalization method has shown that the critical conductivity
subjected to magnetic field is non-universal. Understanding of such
a non-universal behavior gives better insight into physics behind
the superconductor-insulator phase transition phenomena.

The engineering of the conductivity can be also improved by changing
experimentally the value of potential on adjacent sites. This solution
has been recently of growing interest because it allows a simple experimental
realization \cite{Aidelsburger2011,Aidelsburger2013a} and generates
interesting physical phenomena \cite{Delplace2010,Polak2013}. In
particular, tuning the uniaxially periodic potential with a magnetic-flux
quantum per unit cell results in a new possibility of merging cones
simulation in which semi-Dirac point in the Hofstadter spectrum emerges
\cite{Delplace2010}. Recently this possibility has been also exploited
in the context of time-of-light patterns with and without synthetic
magnetic field \cite{Polak2013}.

In this work we present a theory of conductivity valid in a strong
magnetic field in which only intra-Hofstadter-band transitions are taken into account.
In the Mott phase for the square lattice, we describe the optical
behavior using two exemplary values of uniform magnetic field: $f=1/2$
and $f=1/4$ as well as the case with uniaxially staggered potential.
We highlight the fact that the calculation could be straightforwardly
extended to an arbitrary amplitude of the magnetic field and also
gauge degree of freedom. For verification of the results obtained,
a combination of two types of currently available experiments is proposed.
Our method is tested for the the critical value of conductivity on
the Mott insulator - superfluid phase boundary in two dimensions (2D). The proposed
analytical approach correctly reproduces critical conductivity in
the absence of magnetic field $f=0$ (first calculated in \cite{Fisher1990}),
$f=1/2$ (MC numerical solution in Ref. \cite{Cha1994} and analytic
one \cite{Granato1990,Grason2006}\textcolor{black}{{} but using a model
corresponding to BHM}), $f=1/3$ (only numerical solution in Ref.
\cite{Cha1994} based on MC). We have also extended the calculations
of critical conductivity over the range of arbitrary $f=p/q$ which
has not been hitherto reported literature and confirmed its non-universal
behavior upon changes in the amplitude of magnetic field. The influence
of uniaxially staggered potential is also discussed. We compare our
results with presently available numerical and experimental data.

The paper is organized as follows. The BHM in the strong uniform magnetic
field is reviewed in Sec. \ref{sec:Model}. In Sec. \ref{sec:Conductivity-in-magnetic}
we apply our method to study the optical conductivity and its critical
value for different strengths of synthetic magnetic fields, taking
into consideration the effects of uniaxially staggered potential.
In the last section we give a short summary of our results.

\section{Model\label{sec:Model}}

The BHM Hamiltonian in standard notation is given by

\begin{equation}
H=-\sum_{\langle ij\rangle}\tilde{J}_{ij}\left(\hat{b}_{i}^{\dagger}\hat{b}_{j}+\textrm{H.c.}\right)+\frac{U}{2}\sum_{i}\hat{n}_{i}\left(\hat{n}_{i}-1\right)-\mu\sum_{i}\hat{n}_{i}\label{eq:BHM-hamiltonian}
\end{equation}
where $\hat{b}_{i}$ ($\hat{b}_{i}^{\dagger}$) is an operator which
annihilate (create) boson on site $i$ and $\hat{n}_{i}=\hat{b}_{i}^{\dagger}\hat{b}_{i}$.
$U$ and $\mu$ are on-site boson interaction and chemical potential,
respectively. Hopping integral is denoted by $\tilde{J}_{ij}$ and
has a form 
\begin{equation}
\tilde{J}_{ij}=J_{ij}e^{i\frac{e^{*}}{\hbar c}\int_{j}^{i}\mathbf{A}_{0}\cdot d\mathbf{l}}
\end{equation}
with non-zero isotropic factor $J_{ij}=J$ in respect to adjacent
sites. Magnetic field $B$ is introduced by a vector potential $\mathbf{A}_{0}=B(0,x,0)$
which is taken in the Landau gauge. Further in our calculation we
define $Ba^{2}e^{*}/\hbar c=2\pi f$ where $f=Ba^{2}e^{*}/hc=p/q$
is a flux per plaquette ($p$ and $q$ are coprime integers). $f$
depends on the charge of the boson $e^{*}$, lattice spacing $a$,
Planck constant $h$ and speed of light $c$. It is important to stress
that in an optical lattice the quantities like $B$ and $e^{*}$ are
effectively created through tunability of the $p/q$ ratio.

Using the coherent state path integral for the BHM hamiltonian (Eq.
(\ref{eq:BHM-hamiltonian})), the partition function can be written
as follows
\begin{equation}
\mathcal{Z}\left[\mathbf{A}_{0}\right]=\int\mathcal{D}b^{*}\mathcal{D}b\, e^{-\left(S_{0}+S_{1}\left[\mathbf{A}_{0}\right]\right)/\hbar},\label{eq:statistical-sum-0}
\end{equation}
 
\begin{eqnarray}
S_{0} & = & \sum_{i}\int_{_{0}}^{\hbar\beta}d\tau\left\{ b_{i}^{*}(\tau)\hbar\frac{\partial}{\partial\tau}b_{i}(\tau)\right.\label{eq:actionprzed1}\\
 &  & \left.+\frac{U}{2}b_{i}^{*}(\tau)b_{i}^{*}(\tau)b_{i}(\tau)b_{i}(\tau)-\mu b_{i}^{*}(\tau)b_{i}(\tau)\right\} ,\nonumber 
\end{eqnarray}

\begin{equation}
S_{1}\left[\mathbf{A}_{0}\right]=-\sum_{\langle ij\rangle}\int_{_{0}}^{\hbar\beta}d\tau\; J_{ij}e^{i\frac{e^{*}}{\hbar c}\int_{j}^{i}\mathbf{A}_{0}\cdot d\mathbf{l}}\left(b_{i}^{*}(\tau)b_{j}(\tau)+\textrm{c.c.}\right),\label{eq:actionprzed3}
\end{equation}
where the integrals in Eq. (\ref{eq:actionprzed1}) and Eq. (\ref{eq:actionprzed3})
are taken over imaginary time $\tau$ and $\beta=1/k_{B}T$.

To take into account magnetic field in the strong coupling limit of
BHM we follow the same procedure as in Refs. \cite{Sengupta2005a,Sinha2011}
i.e. we perform the double Hubbard-Stratonovich transformation together
with cumulant expansion. In the following, we focus on the Mott phase
and we approximate effective action to second order in the $\mathcal{B}_{\mathbf{k}n}^{q}$
, $\left(\mathcal{B}_{\mathbf{k}n}^{q}\right)^{\dagger}$ fields
\begin{equation}
S^{eff}=\sum_{\mathbf{k}n}\left(\mathcal{B}_{\mathbf{k}n}^{q}\right)^{\dagger}\left[-\hbar G_{0}^{-1}(i\omega_{n})\mathbf{I}+\mathbf{J}^{q}(\mathbf{k})\right]\mathcal{B}_{\mathbf{k}n}^{q},\label{eq:effectiveaction}
\end{equation}
where $\mathbf{I}$ is an identity matrix and
\begin{equation}
\mathcal{B}_{\mathbf{k}n}^{q}=\left(b_{\mathbf{k},n},\; b_{\mathbf{k}+\mathbf{p},n},,\;...,\; b_{\mathbf{k}+(q-1)\mathbf{p},n}\right)^{T},\label{eq:multicomponentfield}
\end{equation}
 with $\mathbf{p}=(2\pi f,\;0)$ and $\omega_{n}=2\pi n/\hbar\beta$
is the Matsubara frequency ($n$ is an integer number). In Eq. (\ref{eq:effectiveaction})
the summation is performed over wave vectors $\mathbf{k}$ within
the first magnetic Brillouin zone where $\left|k_{x}\right|\leqslant\pi/qa$
and $\left|k_{y}\right|\leqslant\pi/a$. $G_{0}(i\omega_{n})$ is
the on-site Green function (i.e. is local where $J=0$) and at the
zero temperature limit it has the form 
\begin{equation}
\frac{1}{\hbar}G_{0}(i\omega_{n})=\frac{n_{0}+1}{i\hbar\omega_{n}+E_{n_{0}}-E_{n_{0}+1}}-\frac{n_{0}}{i\hbar\omega_{n}+E_{n_{0}-1}-E_{n_{0}}}
\end{equation}
with on-site energy $E_{n_{0}}=-\mu n_{0}+Un_{0}(n_{0}-1)/2$. $n_{0}$
is the integer number obtained from the on-site energy minimization
for a given chemical potential $\mu$. The last object in (\ref{eq:effectiveaction})
that requires explanation is $\mathbf{J}^{q}(\mathbf{k})$. It represents
a non-diagonal $q\times q$ matrix similar to that obtained earlier
\cite{Hasegawa1989} 
\begin{equation}
\mathbf{J}^{q}(\mathbf{k})/J=\begin{bmatrix}M_{0} & -e^{ik_{y}a} &  &  & -e^{-ik_{y}a}\\
-e^{-ik_{y}a} & M_{1} & \ddots & 0\\
 & \ddots & \ddots & \ddots\\
 & 0 & \ddots & M_{q-2} & -e^{ik_{y}a}\\
-e^{ik_{y}a} &  &  & -e^{-ik_{y}a} & M_{q-1}
\end{bmatrix},\label{eq:Hasegawa matrix}
\end{equation}
where $M_{\alpha}=-2\cos(k_{x}a+2\pi\alpha f)$. The set of eigenvalues
of the matrix (\ref{eq:Hasegawa matrix}) leads to the Hofstadter
spectrum \cite{Hofstadter1976a}.

It is important to comment here on the notation used above. Although
$b,\; b^{*}$ fields introduced in Eq. (\ref{eq:effectiveaction})
have identical notation as in Eq. (\ref{eq:statistical-sum-0}) they
are not the same. Namely, those fields in which the quadratic effective
action Eq. (\ref{eq:effectiveaction}) is evaluated, were introduced
during the second Hubbard-Stratonovich transformation mentioned above.
However, both kind of fields have the same correlation function \cite{Sengupta2005a}
and for this reason we treat them on an equal footing.

For the effective action in Eq. (\ref{eq:effectiveaction}) we can
find a unitary matrix $U_{q}(\mathbf{k})$ \cite{Saha2010} that diagonalizes
it, i.e.
\begin{equation}
S^{eff}=-\sum_{\mathbf{k}n}\left(\tilde{\mathcal{B}}_{\mathbf{k}n}^{q}\right)^{\dagger}\left[\mathcal{G}^{d}(\mathbf{k},\, i\omega_{n})\right]^{-1}\tilde{\mathcal{B}}_{\mathbf{k}n}^{q}\;,\label{eq:action-eff}
\end{equation}
where $\mathcal{G}^{d}(\mathbf{k},\, i\omega_{n})=U_{q}(\mathbf{k})\left[\hbar G_{0}^{-1}(i\omega_{n})\mathbf{I}-\mathbf{J}^{q}(\mathbf{k})\right]^{-1}U_{q}^{\dagger}(\mathbf{k})$
is an inverse of diagonal Green function and $\tilde{\mathcal{B}}_{\mathbf{k}n}^{q}=U_{q}(\mathbf{k})\mathcal{B}_{\mathbf{k}n}^{q}$.
As shown in Sec. \ref{sec:Conductivity-in-magnetic} this diagonal
form of $\mathcal{G}^{d}(\mathbf{k},\, i\omega_{n})$ has q-bands
which are labeled by $\alpha$ number. 

The above denotations will be useful for further study. In order to
simplify the calculations we also set the lattice constant $a$ and
the reduced Planck constant $\hbar$ to unity.

\section{Conductivity in magnetic field\label{sec:Conductivity-in-magnetic}}

To construct a correct theory of optical conductivity in the strong
magnetic field described by the BHM, we should go beyond the linear
response regime. This situation is more complicated than that previously
studied (e.g., see Ref. \cite{Kampf1993}) due to the presence of
a synthetic magnetic field $\mathbf{A}_{0}$ which significantly modifies
the single-particle spectrum. This modification implies that the commensurability
field effects become important. As a consequence, the amplitude of
hopping term is a complex number (Eq. (\ref{eq:actionprzed3})) and
the boson field is described by more than one component (Eq. (\ref{eq:multicomponentfield})).
To overcome these difficulties in calculation of the transport properties
we assume that such an initial state will be subtly affected by an
additional field $\left|\mathbf{A}\right|\ll\left|\mathbf{A_{0}}\right|$,
which is responsible for generation of optical conductivity in the
linear response regime. Therefore, if we want to study the response
of the system to a strong magnetic field, we simply add to the $\mathbf{A}_{0}$
some small perturbation of this field in the form of a vector potential
$\mathbf{A}$ i.e.

\begin{equation}
\int_{\mathbf{r}_{i}}^{\mathbf{r}_{j}}\mathbf{A}_{0}\cdot d\mathbf{l}\rightarrow\int_{\mathbf{r}_{i}}^{\mathbf{r}_{j}}\mathbf{A}_{0}\cdot d\mathbf{l}+\int_{\mathbf{r}_{i}}^{\mathbf{r}_{j}}\mathbf{A}\cdot d\mathbf{l}
\end{equation}
and use the linear response theory as a starting point with respect
to the quantity $\mathbf{A}$. This leads to the well known expression
for the optical conductivity (e.g., see Refs. \cite{Wu2006,Kampf1993,Cha1991})
\begin{equation}
\sigma_{xx}^{\mathbf{A}_{0}}(i\omega)=-\left.\frac{1}{N\omega}\sum_{ij}\int_{0}^{\beta}d\tau e^{i\omega\tau}\frac{\delta^{2}\ln\mathcal{Z}\left[\mathbf{A}'\right]}{\delta A_{i}^{x}(\tau)\delta A_{j}^{x}(0)}\right|_{\mathbf{A}=0},\label{eq:conduct_derivative}
\end{equation}
but in our case the conductivity depends upon vector potential $\mathbf{A}_{0}$
and consequently on magnetic field and in this dependence the commensurability
effects are included. 

Performing the calculations in Eq. (\ref{eq:conduct_derivative})
we first go to the wave vector representation which allows proper
incorporation of the perturbing vector potential $\mathbf{A}$, through
the substitution $k_{x}\rightarrow k_{x}+\frac{e^{*}}{\hbar c}\mathbf{A}$
(we assume that $\mathbf{A}$ does not depend on position) and then
we evaluate the derivatives in Eq. (\ref{eq:conduct_derivative})
getting
\begin{equation}
\sigma_{xx}^{\mathbf{A}_{0}}(i\omega)=-\frac{\left(e^{*}\right)^{2}}{\omega}\left\langle e_{kin}^{x}\right\rangle +\frac{1}{\omega}\Pi_{xx}(i\omega)\;,\label{eq:general_form_of_conductivity}
\end{equation}
where functional averages are taken with the partition function from
Eq. (\ref{eq:statistical-sum-0}). $e_{kin}^{x}$ and $\Pi_{xx}(\omega)$
are a kinetic and current-current correlation function, respectively
(we set $c=1$). With the effective action from Eq. (\ref{eq:action-eff})
we can evaluate Eq. (\ref{eq:general_form_of_conductivity}) to the
form
\begin{equation}
\Pi_{xx}(i\omega)=-\frac{1}{N}\sum_{\mathbf{kk'}}\int_{0}^{\beta}d\tau\; e^{i\omega\tau}\left\langle j_{\mathbf{k}}^{x}(\tau)j_{\mathbf{k'}}^{x}(0)\right\rangle ,\label{eq:paramagnetic}
\end{equation}
\begin{equation}
e_{kin}^{x}=-\frac{1}{N}\sum_{\mathbf{k}}\sum_{\alpha=0}^{q-1}\left[\partial_{k_{x}}^{2}\epsilon_{q}^{\alpha}(\mathbf{k};p)\right]b_{\mathbf{k}+\alpha\mathbf{p}}^{*}(0)b_{\mathbf{k}+\alpha\mathbf{p}}(0),\label{eq:diamagnetic}
\end{equation}
\begin{equation}
j_{\mathbf{k}}^{x}(\tau)=e^{*}\sum_{\alpha=0}^{q-1}\left[\partial_{k_{x}}\epsilon_{q}^{\alpha}(\mathbf{k};p)\right]b_{\mathbf{k}+\alpha\mathbf{p}}^{*}(\tau)b_{\mathbf{k}+\alpha\mathbf{p}}(\tau),\label{eq:current}
\end{equation}
where $\epsilon_{q}^{\alpha}(\mathbf{k};p)$ is an eigenvalue of $J_{q}(\mathbf{k})$
matrix defined in Eq. (\ref{eq:Hasegawa matrix}) and the current
is given by Eq. (\ref{eq:current}). Eq. (\ref{eq:general_form_of_conductivity})
is the Kubo formula valid in strong magnetic field (MKF)  in which only intra-Hofstadter-band transitions are taken into account. This method,
in contrast to that presented in Ref. \cite{VanOtterloA1993} (see
also Refs. \cite{Wagenblast1996,Wagenblast1997,Dalidovich2001b,Dalidovich2002,Wu2006}),
takes into account the commensurability effects of the magnetic field
and covers the whole range of $\mathbf{k}$ and $\omega$ dependence. 

For the simplest case, if we take the usual square lattice dispersion
relation in the absence of magnetic field $\epsilon_{1}^{\alpha}(\mathbf{k};0)=-2J\left(\cos k_{x}+\cos k_{y}\right)$,
one gets from Eq. (\ref{eq:general_form_of_conductivity}) 
\begin{eqnarray}
\sigma_{xx}^{\mathbf{A}_{0}}(i\omega) & = & \frac{\left(e^{*}\right)^{2}J}{\omega}\frac{1}{N}\sum_{\mathbf{k}}2\cos k_{x}\left\langle b_{\mathbf{k}}^{*}(0)b_{\mathbf{k}}(0)\right\rangle \label{eq:cond3}\\
 &  & -\frac{4\left(e^{*}\right)^{2}J^{2}}{\omega}\frac{1}{N}\sum_{\mathbf{k},\mathbf{k}'}\int_{0}^{\beta}d\tau\,\mathrm{e}^{i\omega\tau}\times\nonumber \\
 &  & \sin(k_{x})\sin(k'_{x})\langle b_{\mathbf{k}}^{*}(\tau)b_{\mathbf{k}}(\tau)b_{\mathbf{k}'}^{*}(0)b_{\mathbf{k}'}(0)\rangle\;,\nonumber 
\end{eqnarray}
which recovers a well known result \cite{Kampf1993}.

\begin{figure}[th]
\includegraphics[scale=0.47]{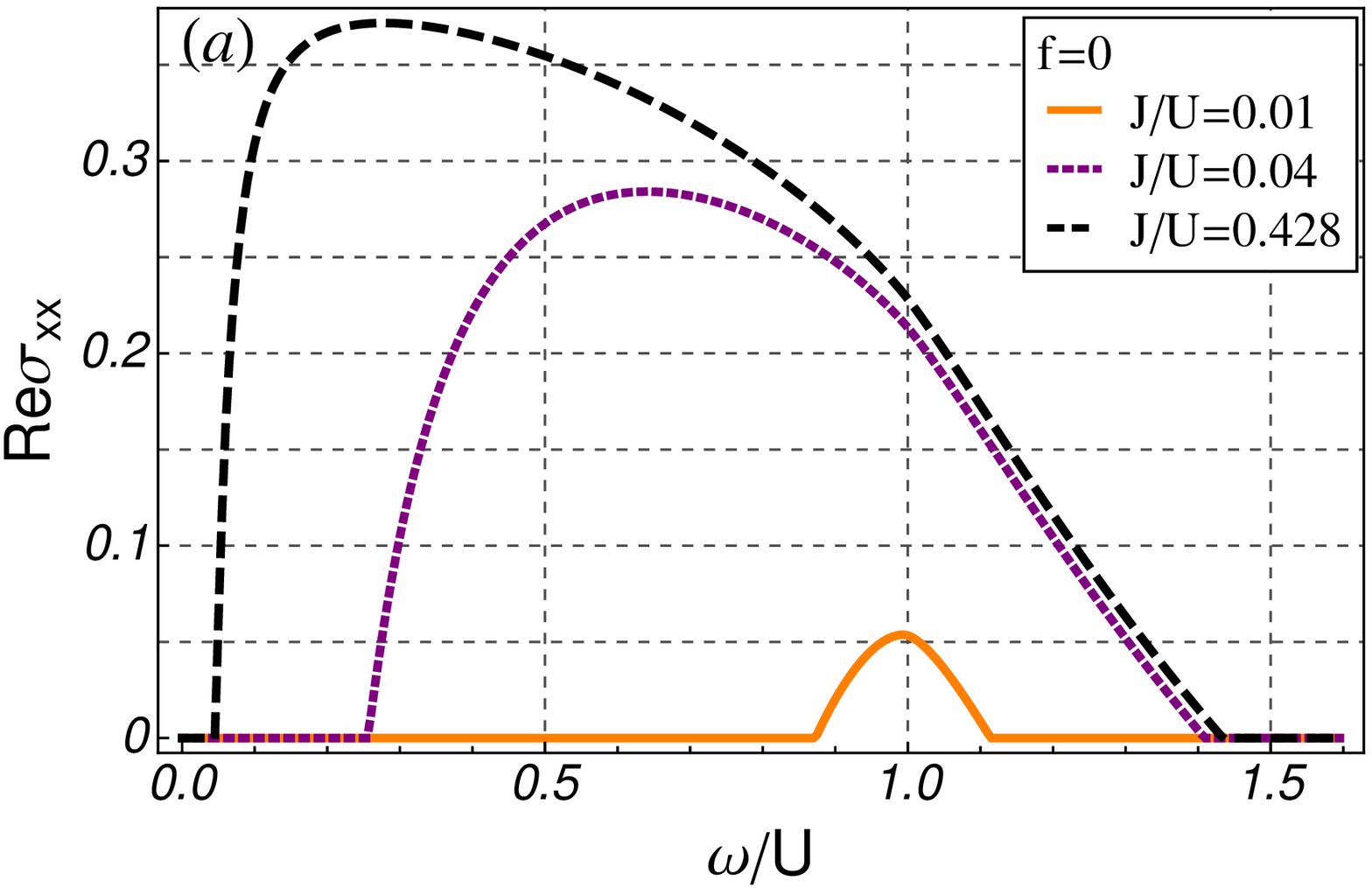}

\includegraphics[scale=0.47]{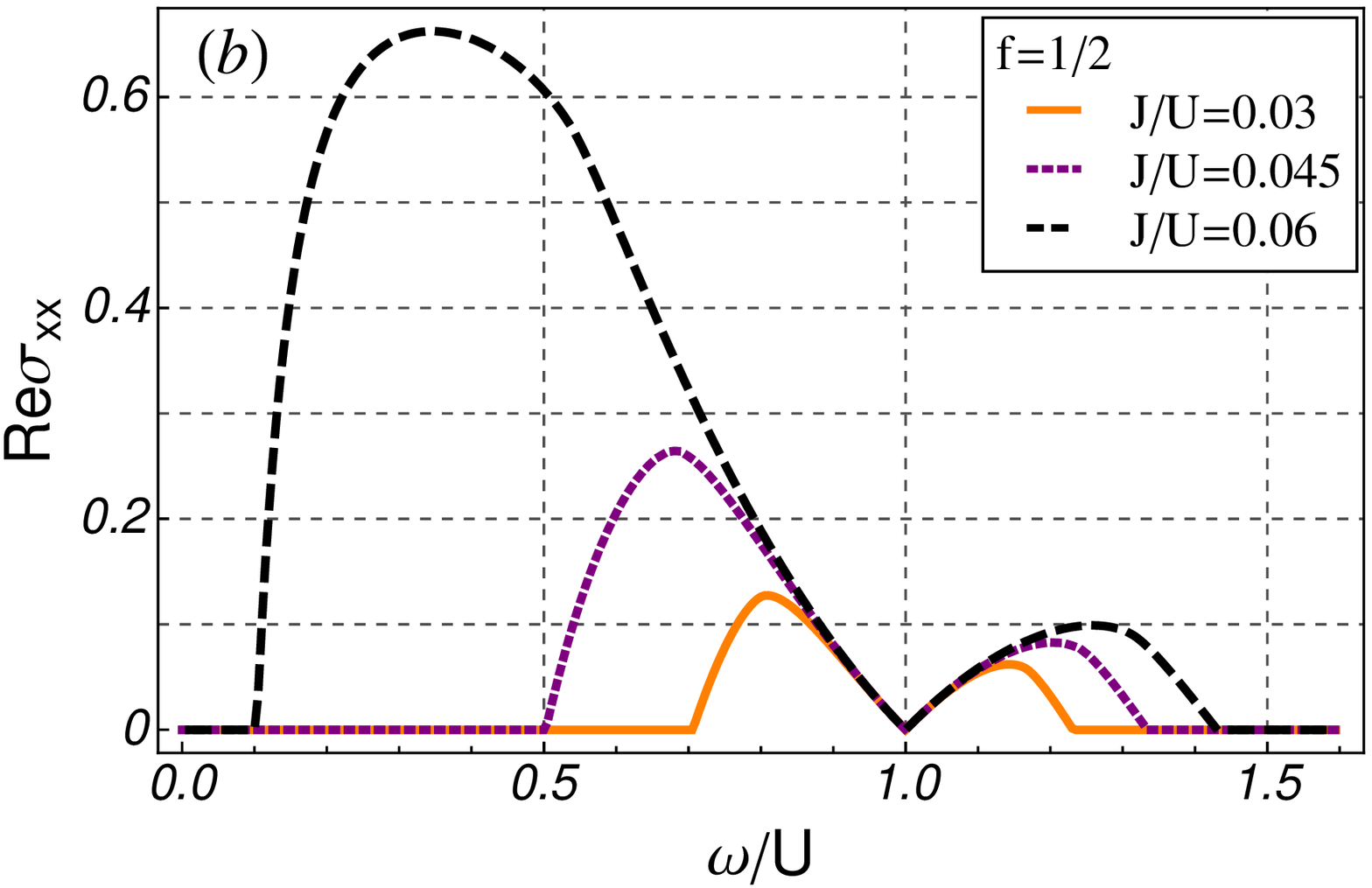}

\includegraphics[scale=0.47]{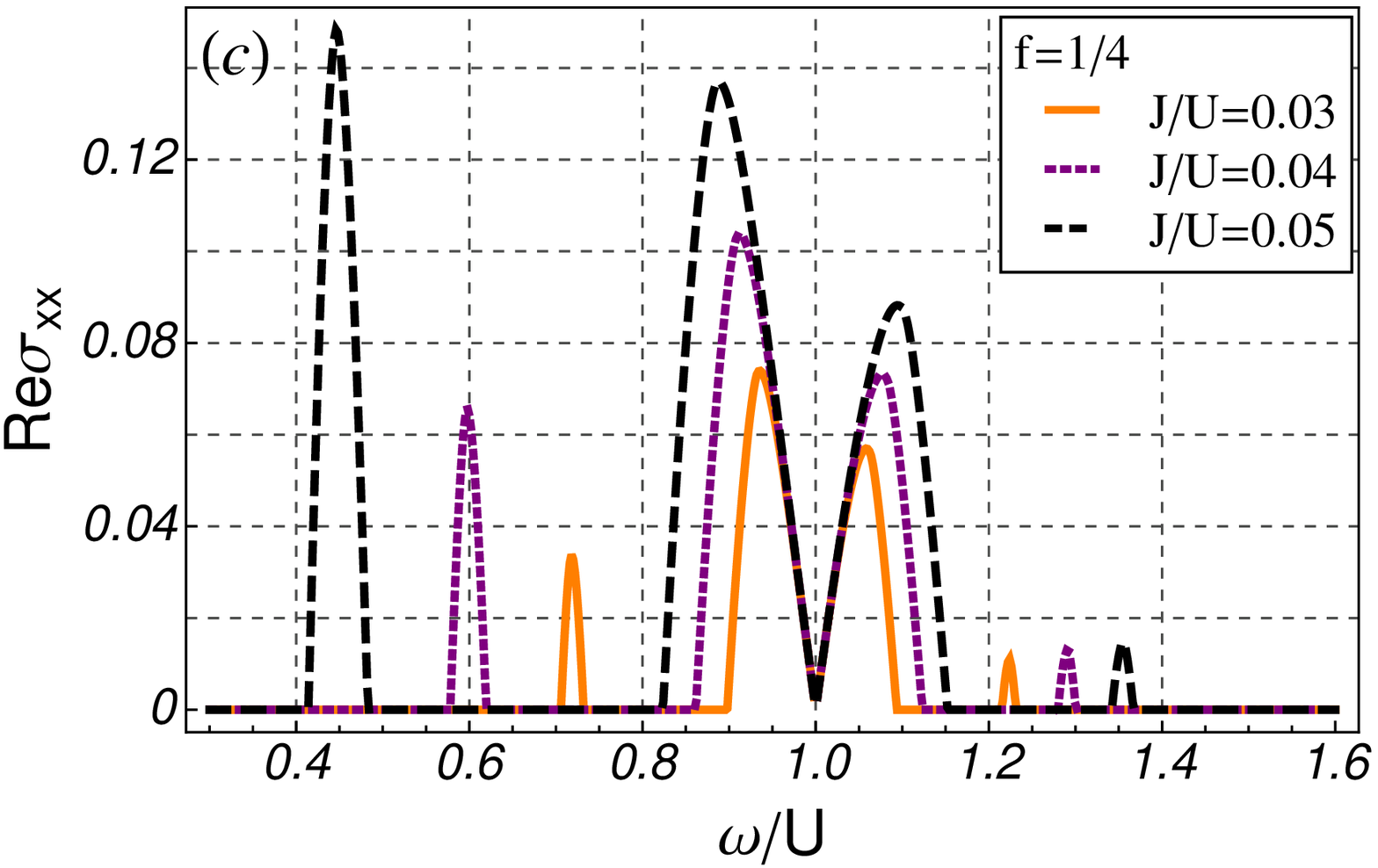}

\caption{Real part of optical conductivity in the two-dimensional square lattice
in Mott phase (first lobe) is sketched in $\sigma_{Q}$ units ($\sigma_{Q}=\left(e^{*}\right)^{2}/h$
is a quantum conductance). From the top - $f=0$ (a), $f=1/2$ (b),
$f=1/4$ (c). Conductivity is plotted at zero temperature for different
value of $J/U$ within the first lobe.\label{fig:Optical-conductivity-in}}
\end{figure}

\subsection{Optical conductivity in the BHM\label{sub:Optical-conductivity-1-2-4}}

\subsubsection{The uniform field\label{sub:Landau-gauge}}

Now we are interested in the OC in the Mott insulator phase. Using
the fact that within the action (\ref{eq:action-eff}) the four point
correlation function in Eq (\ref{eq:paramagnetic}) is factorized,
we can rewrite Eq (\ref{eq:general_form_of_conductivity}) to the
form
\begin{eqnarray}
\sigma_{xx}^{\mathbf{A}_{0}}(i\omega) & = & -\frac{\left(e^{*}\right)^{2}J}{\omega}\frac{1}{\beta N}\sum_{\mathbf{k}n}\sum_{\alpha=0}^{q-1}\partial_{k_{x}}^{2}\epsilon_{q}^{\alpha}(\mathbf{k};p)\mathcal{G}_{\alpha\alpha}^{d}(\mathbf{k},i\omega_{n})\nonumber \\
 &  & -\frac{\left(e^{*}\right)^{2}}{\omega}\frac{1}{\beta N}\sum_{\mathbf{k}n}\sum_{\alpha=0}^{q-1}\times\label{eq:cond_A final form}\\
 &  & \left[\partial_{k_{x}}\epsilon_{q}^{\alpha}(\mathbf{k};p)\right]^{2}\mathcal{G}_{\alpha\alpha}^{d}(\mathbf{k},i\omega_{n})\mathcal{G}_{\alpha\alpha}^{d}(\mathbf{k},i\omega_{n}+i\omega).\nonumber 
\end{eqnarray}
Mott insulator Green's function is
\begin{eqnarray}
\mathcal{G}_{\alpha\alpha}^{d}(\mathbf{k},i\omega_{n}) & = & \frac{G_{0}(i\omega_{n})}{1-\epsilon_{q}^{\alpha}(\mathbf{k};p)G_{0}(i\omega_{n})}\\
 & = & \frac{z_{q}^{\alpha}(\mathbf{k};p)}{i\omega_{n}-E_{q}^{\alpha+}(\mathbf{k};p)}+\frac{1-z_{q}^{\alpha}(\mathbf{k};p)}{i\omega_{n}-E_{q}^{\alpha-}(\mathbf{k};p)}\,,\nonumber 
\end{eqnarray}
where the weight and dispersion of quasiparticles have the form
\begin{equation}
z_{q}^{\alpha}(\mathbf{k};p)=\frac{E_{q}^{\alpha+}(\mathbf{k};p)+\mu+U}{E_{q}^{\alpha+}(\mathbf{k};p)-E_{q}^{\alpha-}(\mathbf{k};p)}\;,
\end{equation}
\begin{equation}
E_{q}^{\alpha\pm}(\mathbf{k};p)=\frac{\epsilon_{q}^{\alpha}(\mathbf{k};p)}{2}-\mu+U\left(n_{0}-\frac{1}{2}\right)\pm\frac{1}{2}\Delta_{q}^{\alpha}(\mathbf{k};p)
\end{equation}
\begin{equation}
\Delta_{q}^{\alpha}(\mathbf{k};p)=\sqrt{\left[\epsilon_{q}^{\alpha}(\mathbf{k};p)\right]^{2}+4\epsilon_{q}^{\alpha}(\mathbf{k};p)U\left(n_{0}+\frac{1}{2}\right)+U^{2}}\label{eq:gap}
\end{equation}

In the following we are interested in the real and regular part of
optical conductivity given by 
\begin{eqnarray}
 &  & \textrm{Re}\sigma_{xx}^{\mathbf{A}_{0}}(\omega)=\left(e^{*}\right)^{2}\frac{2\pi}{N}\sum_{\alpha=0}^{q-1}\sum_{\mathbf{k}}\left(\partial_{k_{x}}\epsilon_{q}^{\alpha}(\mathbf{k};p)\right)^{2}\label{eq:optical_reg}\\
 &  & \times\left\{ n_{B}\left[E_{q}^{\alpha-}(\mathbf{k};p)\right]-n_{B}\left[E_{q}^{\alpha+}(\mathbf{k};p)\right]\right\} \nonumber \\
 &  & \times\left[1-z_{q}^{\alpha}(\mathbf{k};p)\right]z_{q}^{\alpha}(\mathbf{k};p)\delta\left(\omega^{2}-\left(\Delta_{q}^{\alpha}(\mathbf{k};p)\right)^{2}\right)\;,\nonumber 
\end{eqnarray}
which has been obtained by evaluating Matsubara sum in Eq. (\ref{eq:cond_A final form})
and performing standard analytical continuation $i\omega\rightarrow\omega+i\delta$
\footnote{The same result for the conductivity in the Mott phase can be obtained
by using an effective quadratic action coming from single Hubbard-Stratonovich
transformation of the hopping term. {[}A. S. Sajna, unpublished{]}%
}. The symbol $\delta(x)$ is a Dirac delta function and $n_{B}(x)$
is Bose-Einstein distribution function. Using density of states and
taking zero-temperature limit, equation (\ref{eq:optical_reg}) can
be rewritten as follows 
\begin{eqnarray}
\textrm{Re}\sigma_{xx}^{\mathbf{A}_{0}}(\omega)_{\textrm{}} & = & 2\pi^{2}\sigma_{Q}\sum_{\alpha}\sum_{s=\{+,-\}}\Xi_{q}^{\alpha}\left[u^{s}(\omega);p\right]\;,
\end{eqnarray}
\begin{eqnarray}
\Xi_{q}^{\alpha}\left[v;p\right] & = & \rho_{q}^{\alpha}(v;p)\frac{J\left[z_{q}^{\alpha}(v;p)-1\right]z_{q}^{\alpha}(v;p)}{U\sqrt{4n_{0}\left(n_{0}+1\right)+\left(\omega/U\right)^{2}}}\;,
\end{eqnarray}
\begin{eqnarray}
u^{\pm}(\omega) & = & \frac{U}{J}\left(2n_{0}+1\right)\left(1\mp\sqrt{1-\frac{1-\left(\omega/U\right)^{2}}{\left(2n_{0}+1\right)^{2}}}\right)\;,
\end{eqnarray}
where the density of states is given by
\begin{equation}
\rho_{q}^{\alpha}(v;p)=\frac{1}{N}\sum_{\mathbf{k}}\left[\partial_{k_{x}}\epsilon_{q}^{\alpha}(\mathbf{k};p)\right]^{2}\delta\left(v-\epsilon_{q}^{\alpha}(\mathbf{k};p)\right)\;.\label{eq:DOS-cond}
\end{equation}
For magnetic fields considered here (i.e. $f=0,\;1/2,\ 1/4$) the
exact calculations of density of states in 2D was possible in terms
of complete elliptic integrals (see Appendix \ref{sub:appendix-uniform}).

\begin{figure}[tb]
\includegraphics[scale=0.245]{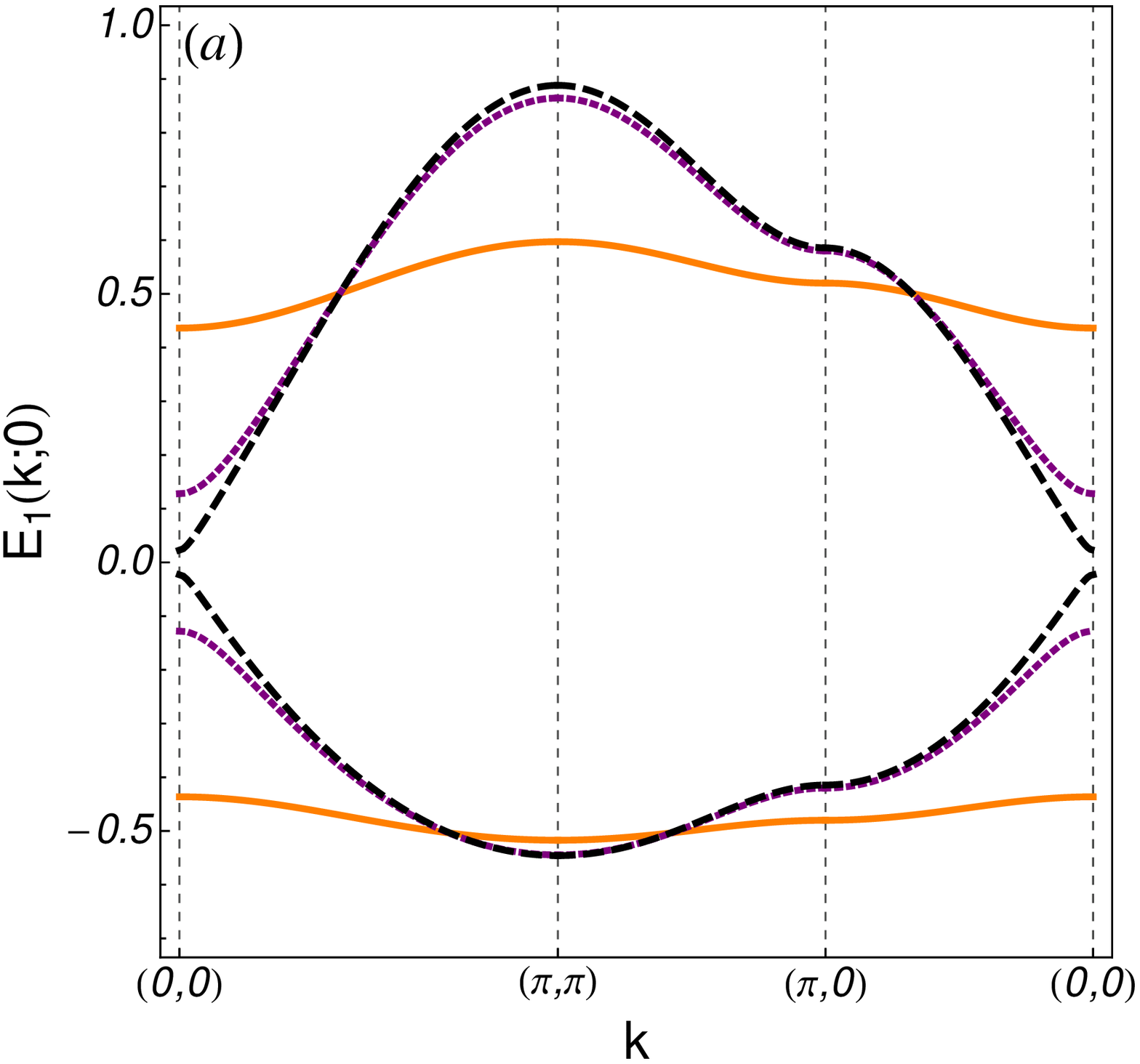}\includegraphics[scale=0.245]{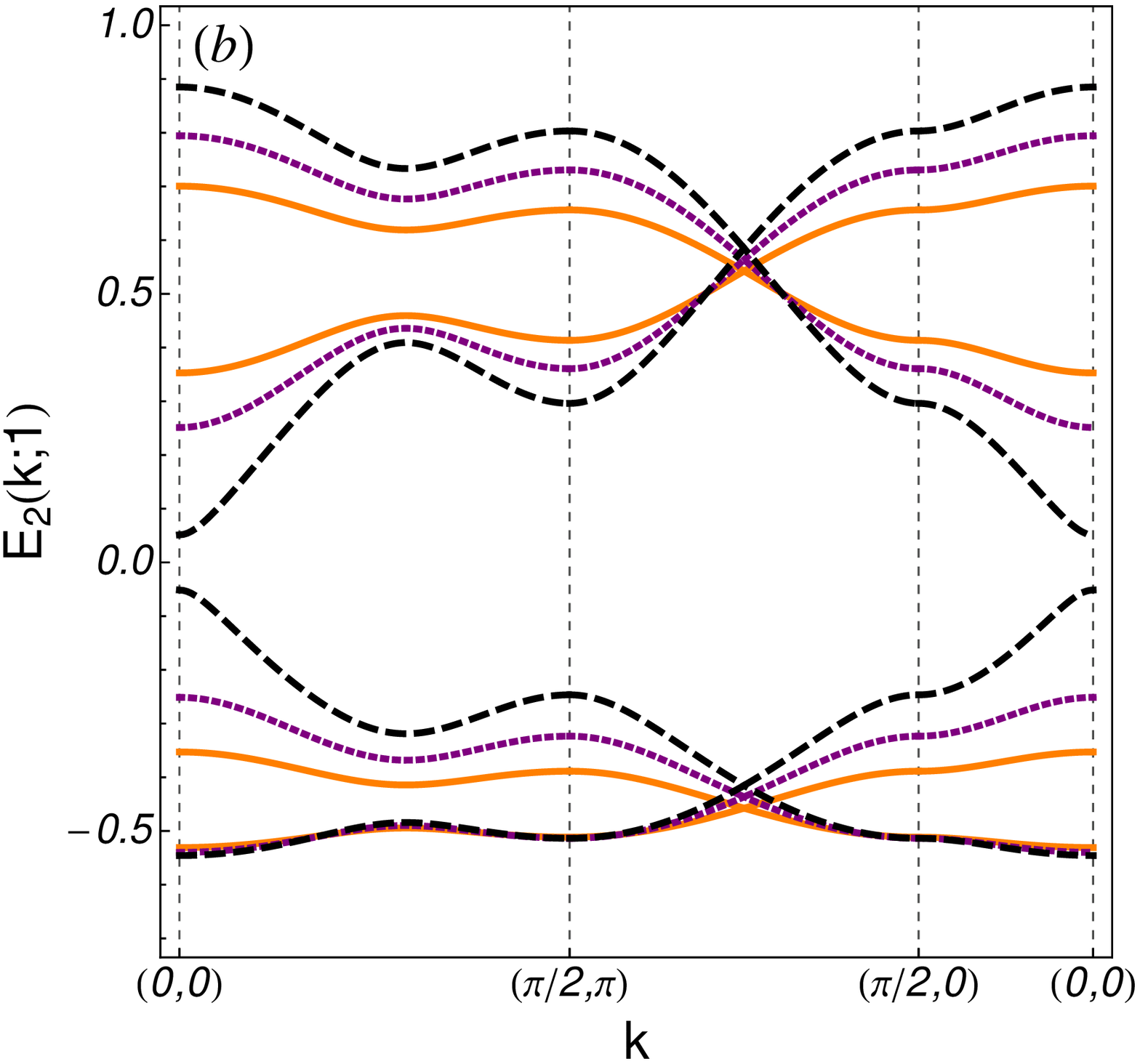}

\includegraphics[scale=0.245]{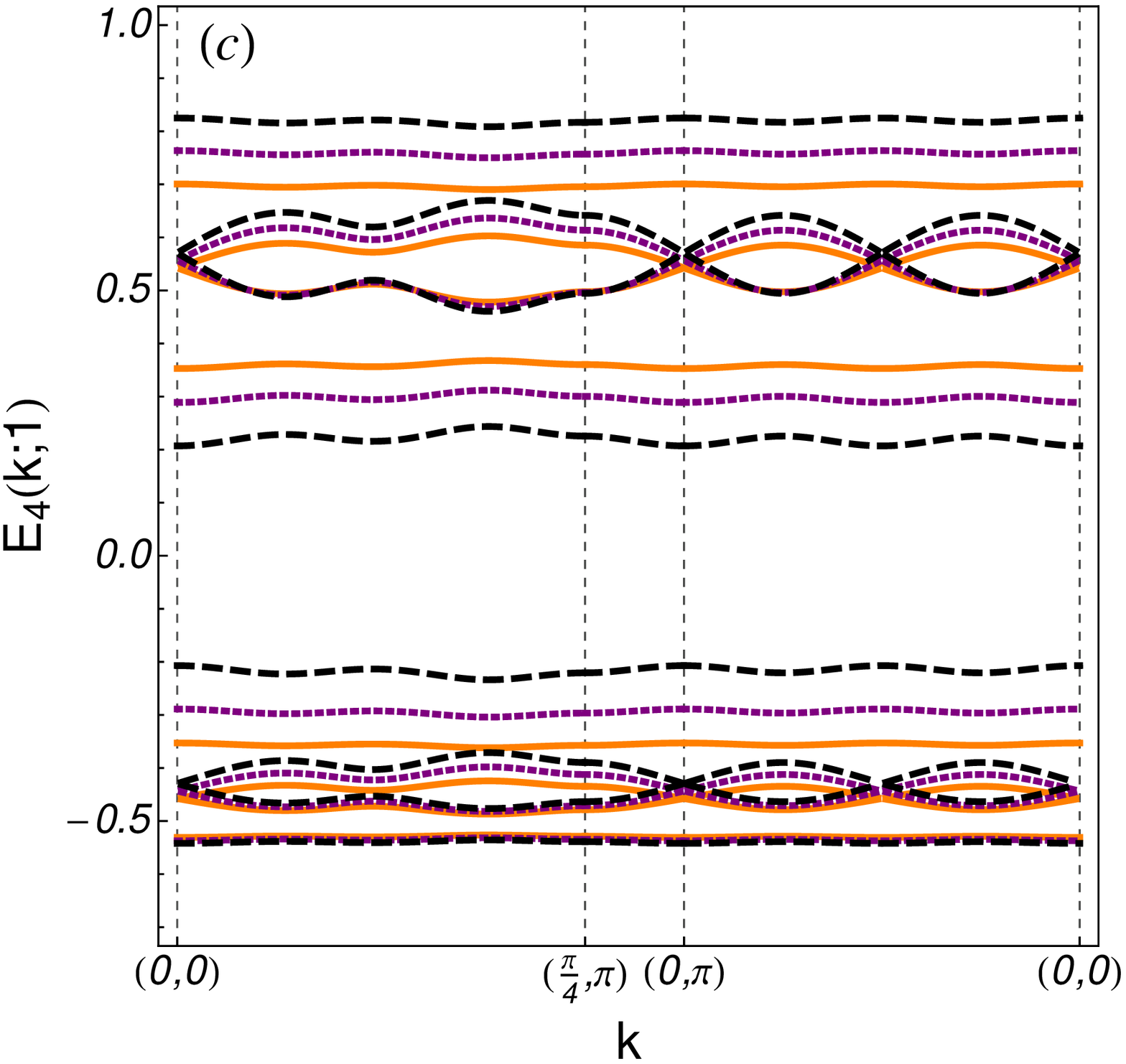}\includegraphics[scale=0.245]{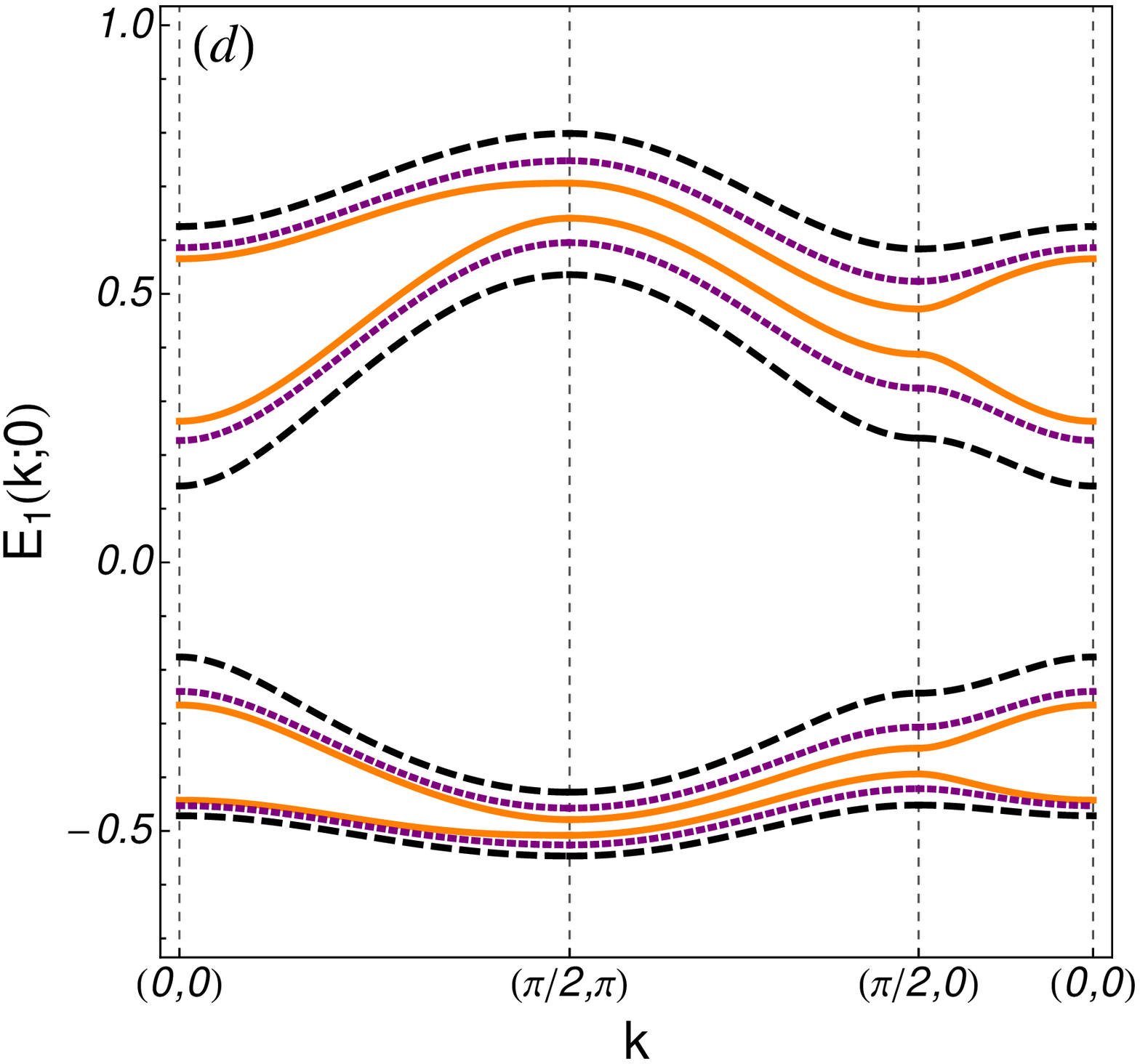}

\includegraphics[scale=0.245]{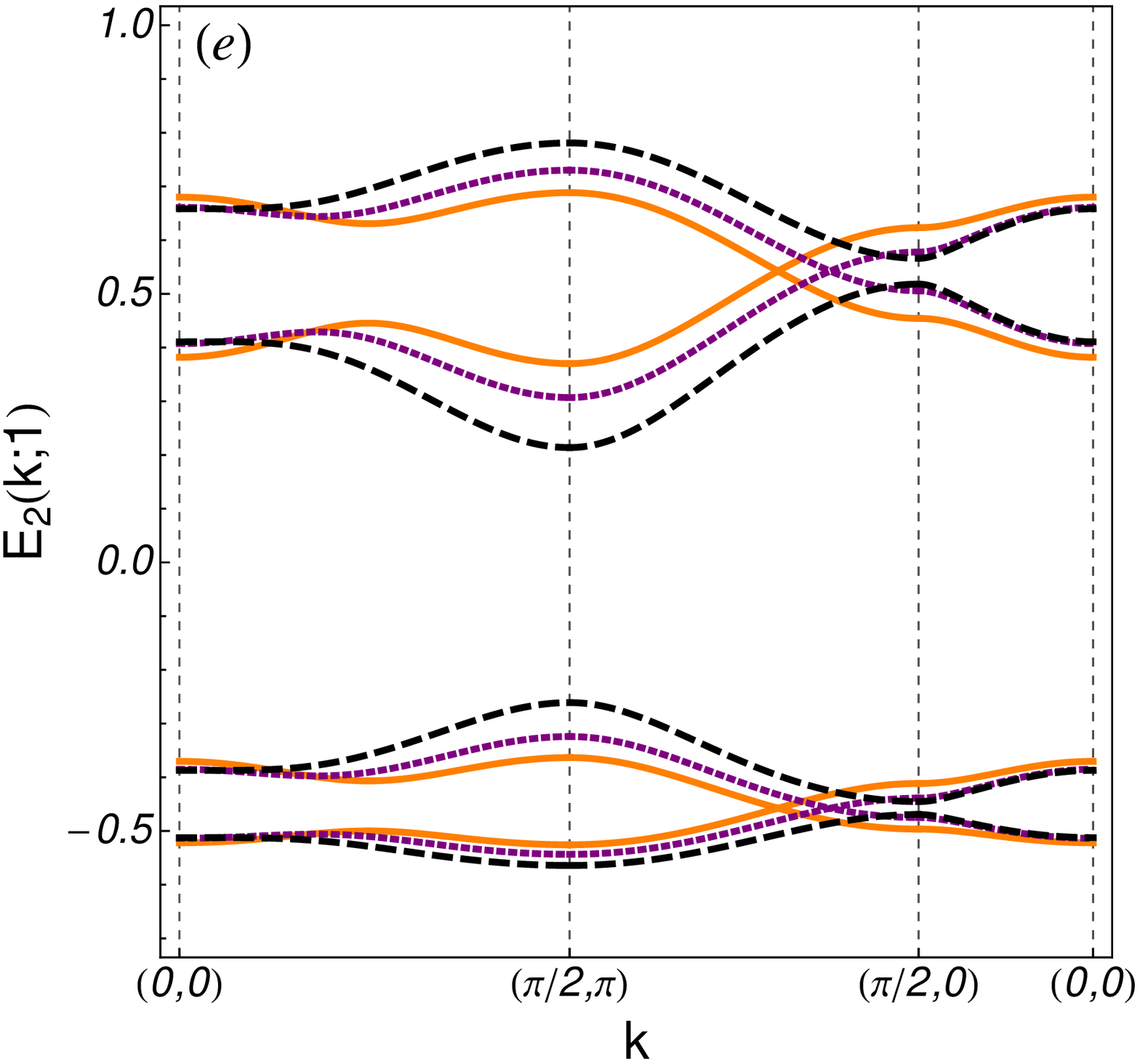}

\caption{The wave vector $\mathbf{k}$ dependence of the quasi-particle energy
dispersions $E_{q}^{\alpha\pm}(\mathbf{k};p)$ for different amplitudes
of uniform magnetic field $f$ and staggered potential $\Delta$ in
the first magnetic Brillouin zone. Plots $a$, $b$, $c$ correspond
to the Figure \ref{fig:Optical-conductivity-in} $a$, $b$, $c$,
respectively. Additionally, plots $d$ and $e$ correspond to Fig.
\ref{fig:f-0_with_Delta} and Fig. \ref{fig:Frist-(second)-line},
respectively. This correspondence is revealed when choosing the same
physical parameters for the same plotting style in the relevant figures.\label{fig:quasi}}

\end{figure}

In Fig. \ref{fig:Optical-conductivity-in} we show the real part of
OC at zero temperature in the Mott phase for different values of synthetic
magnetic field. It is worth noting that its behavior reflects the
tight binding dispersion of the lattice. As expected the OC is gradually
broadened at the cost of vanishing gap when $J/U$ increases. Interestingly,
for $f=1/4$, the contribution of the lowest frequency peak becomes
much more significant when the $J/U$ ratio is tuned up.

For $\omega=U$, we observe that the existence of conductivity at
this point directly depends on the spectrum weight value of the tight
binding dispersion $\epsilon_{q}^{\alpha}(\mathbf{k};p)$. In particular,
if the spectrum weight of $\epsilon_{q}^{\alpha}(\mathbf{k};p)$ at
the center of the band is zero the conductivity also disappears. Such
a situation for $\omega=U$ is satisfied when $f=1/2,\;1/4$ (even
$q$) but for odd $q$ we should observe metallic behavior. This special
behavior of OC at this point is directly related to the Dirac cones
appearing in quasi-particle energy dispersion $E_{q}^{\alpha\pm}(\mathbf{k};p)$.
The corresponding dispersions are plotted in Fig. 2 $a-c$ for the
relevant set of parameters. In the Sec. \ref{sub:Connection-to-experiment}
we suggest an experiment to check this conjecture explicitly, because
the quantities like the on-site interaction strength $U$ and boson
hopping amplitude $J$ are fully controllable parameters in ultra-cold
gases loaded on optical lattice.

\begin{figure}[tb]
\includegraphics[scale=0.47]{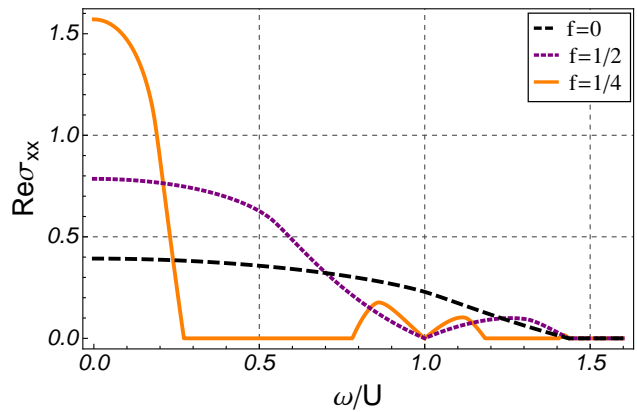}

\caption{Optical conductivity for different values of magnetic field. The parameters
$\mu/U$ and $J/U$ are chosen on the phase boundary where Mott insulator
- superfluid phase transition takes a place. The plot shows that the
critical value of conductivity for the $f=1/2$ ($1/4$) when $\omega\rightarrow0$
is two (four) times higher than in the case when magnetic field is
absent. $\textrm{Re}\sigma_{xx}$ was plotted in $\sigma_{Q}$ units.
\label{fig:critical_cond_from_OC}}

\end{figure}

Within the above framework we can also probe the critical value of
conductivity at the tip of the lobe (for determination of the critical
value of $\mu/U$ and $J/U$ see Ref. \cite{Sinha2011}). For this
range of parameters the OC is as shown in Fig. \ref{fig:critical_cond_from_OC}.
The critical value of conductivity for $f=1/2$ ($1/4$) when $\omega\rightarrow0$
is two (four) times higher than the value with magnetic field absent.
These results are in agreement with those derived in Sec. \ref{sub:Critical-conductivity}.

\begin{figure*}[th]
\includegraphics[scale=0.55]{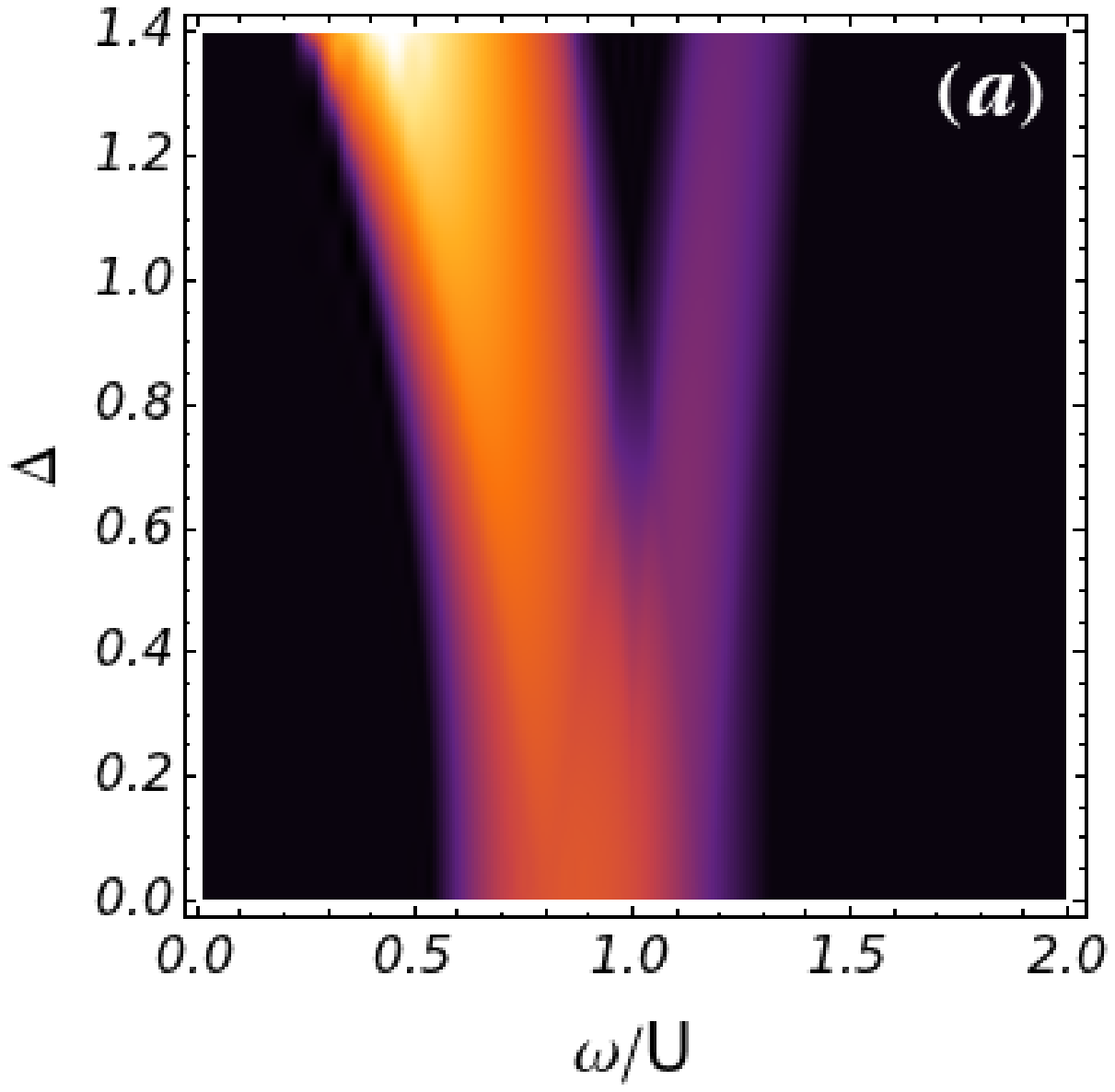}\includegraphics[scale=0.57]{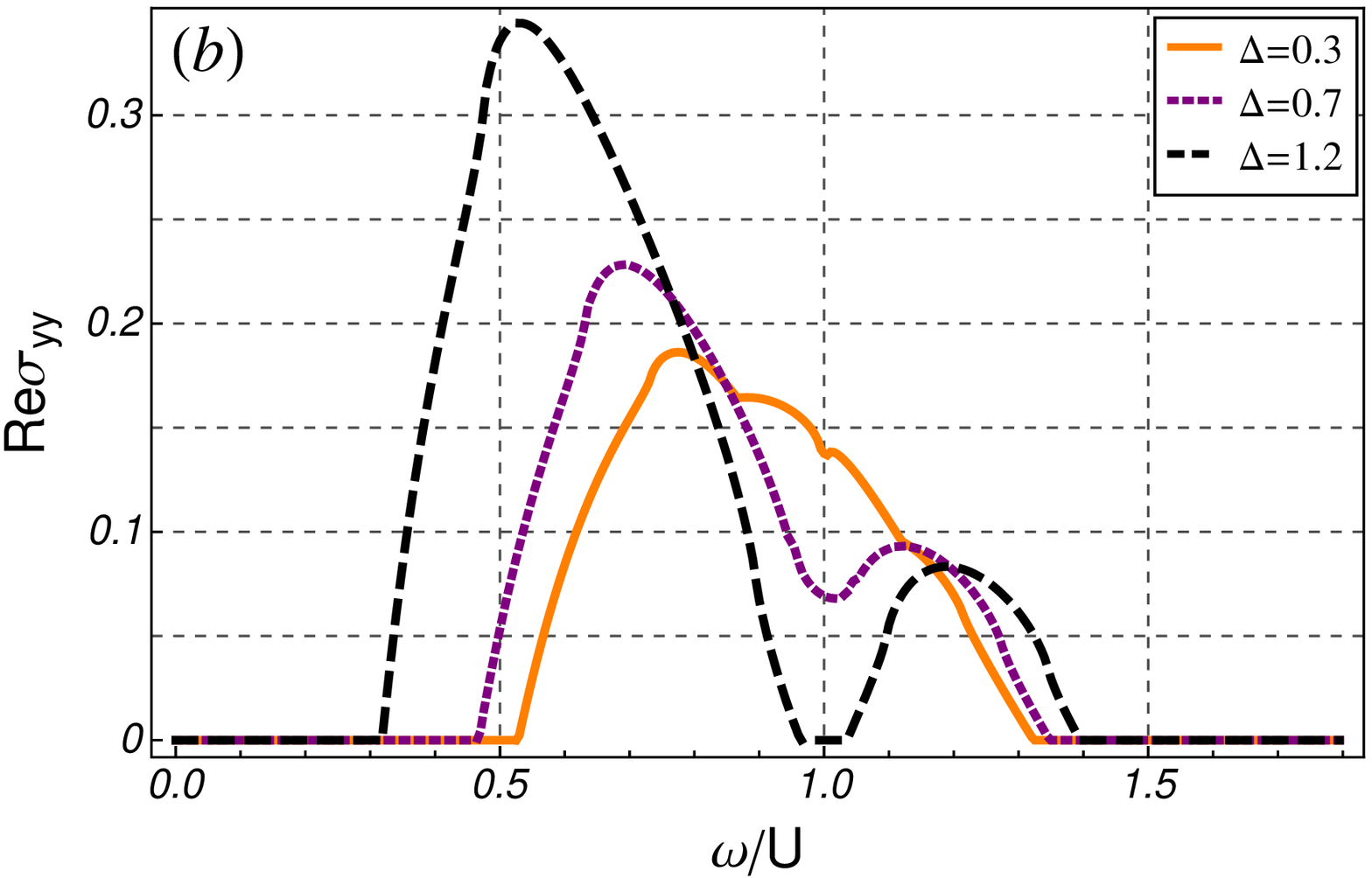}

\includegraphics[scale=0.55]{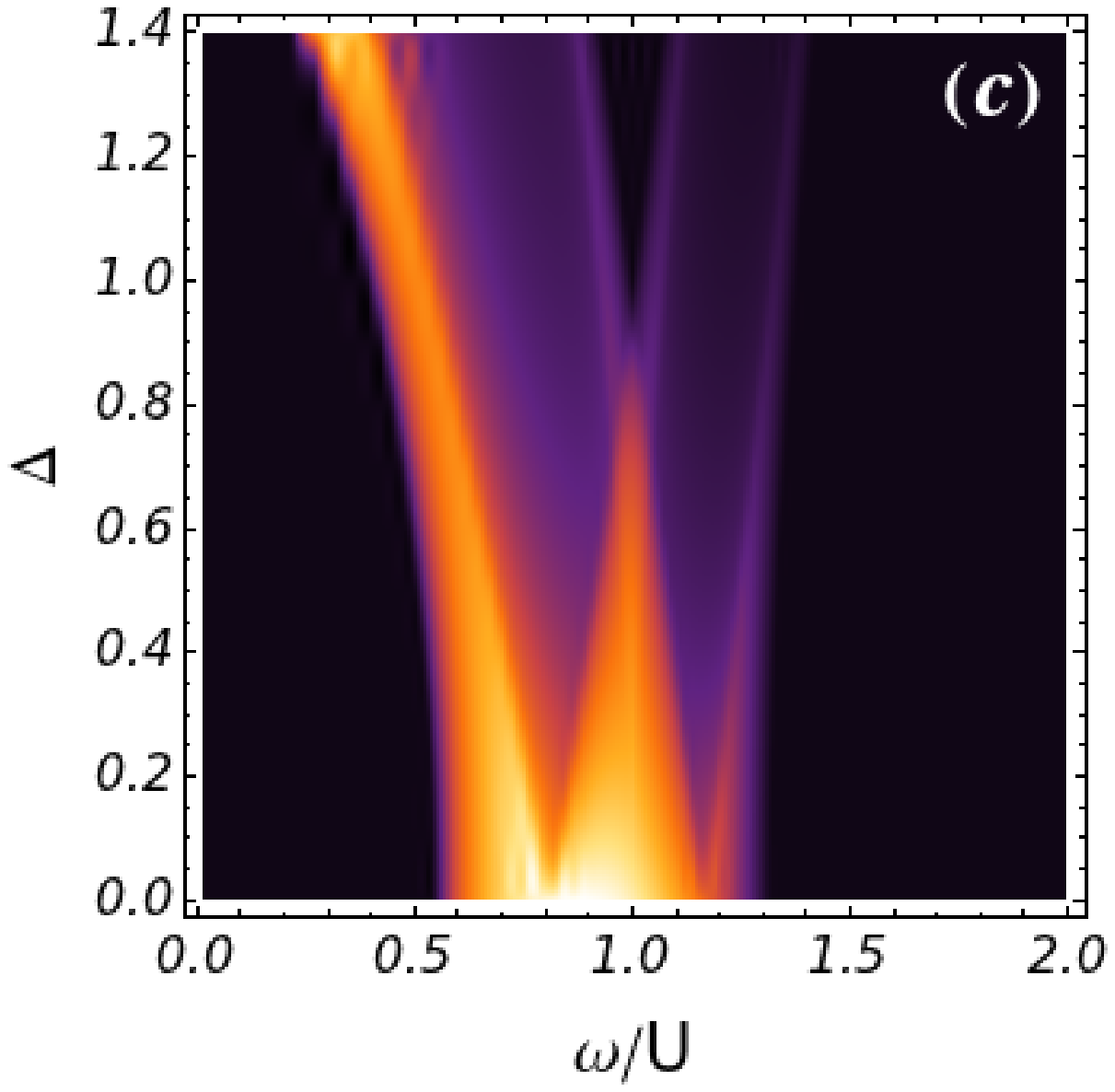}\includegraphics[scale=0.57]{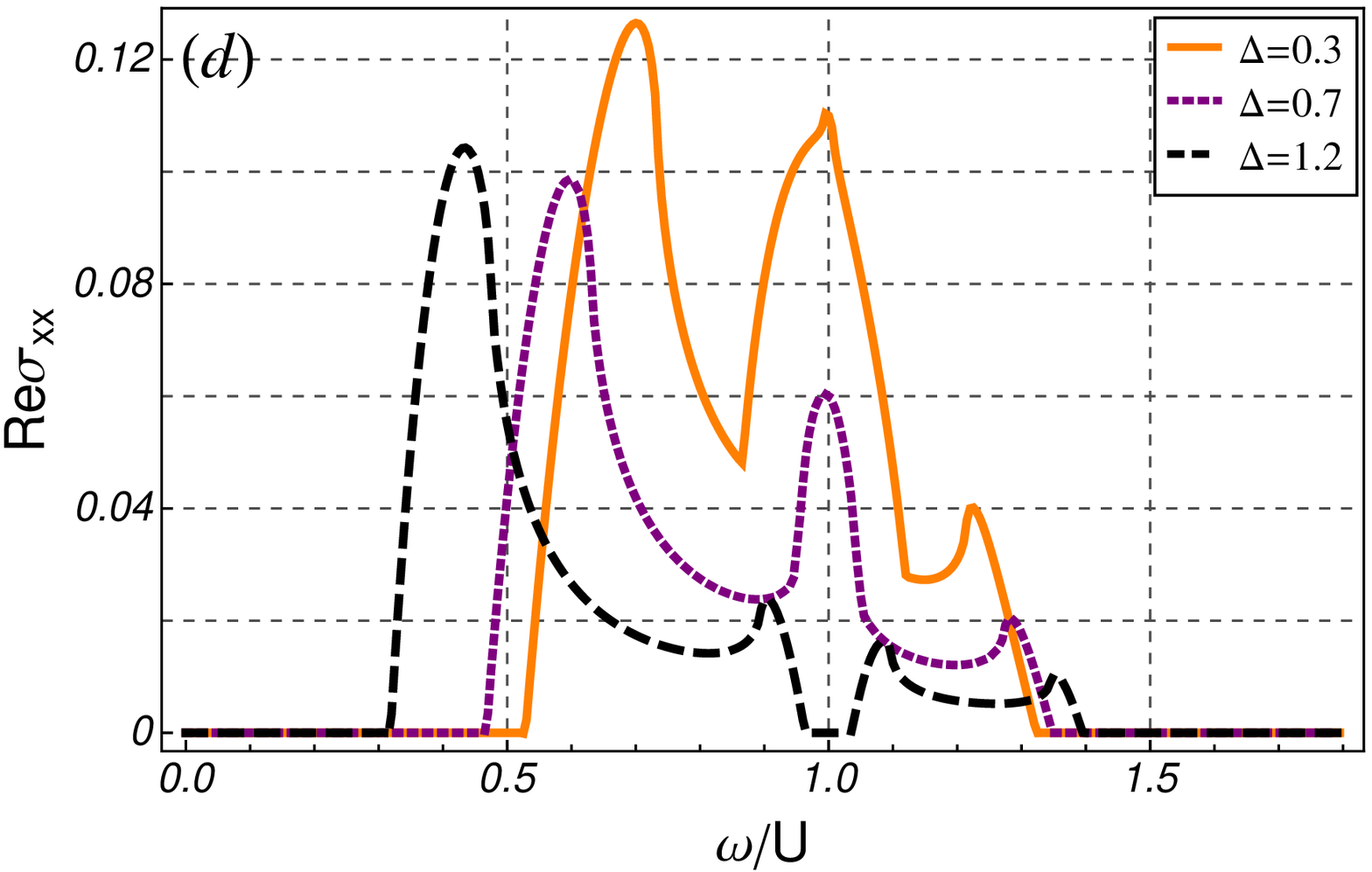}

\caption{First (second) panel shows $yy$ ($xx$) component of the real part
of the optical conductivity for different values of $\Delta$ in the
absence of magnetic field. On the density sketch the lighter color
indicates higher amplitude of the conductivity (black area correspond
to the insulator behavior). All zero-temperature plots show the situation
where the uniaxially staggered field is subjected to the $x$ axis.
The ratio of hopping amplitude and on-site interaction energy is $J/U=0.03$
and the conductivity was plotted in $\sigma_{Q}$ units.\label{fig:f-0_with_Delta}}
\end{figure*}

\subsubsection{The uniform field with staggered potential\label{sub:Staggered-potential}}

In the following we use the proposed framework of OC (see Eq. (\ref{eq:general_form_of_conductivity}))
to study the transport phenomena under the influence of uniaxially
staggered potential. The situations with the synthetic magnetic field
absent ($f=0$) and with its value described by half flux per plaquette
($f=1/2$) are considered. The latter special case is a subject of
current interest \cite{Delplace2010}. In both cases of OC i.e. $f=0$
and $f=1/2$ the staggered potential is controlled by the parameter
$\Delta$ \cite{Polak2013}, which is included in the Hamiltonian
(\ref{eq:BHM-hamiltonian}) by an additional term in the form
\begin{equation}
\sum_{i_{x}i_{y}}(-1)^{i_{x}}\tilde{\Delta}\hat{b}_{i_{x}i_{y}}^{\dagger}\hat{b}_{i_{x}i_{y}}\;,
\end{equation}
where $\Delta=\tilde{\Delta}/2J$ and $i_{x},\; i_{y}$ enumerate
the positions of lattice sites along the $x$ and $y$ axis, respectively. 

\begin{figure*}[th]
\includegraphics[scale=0.55]{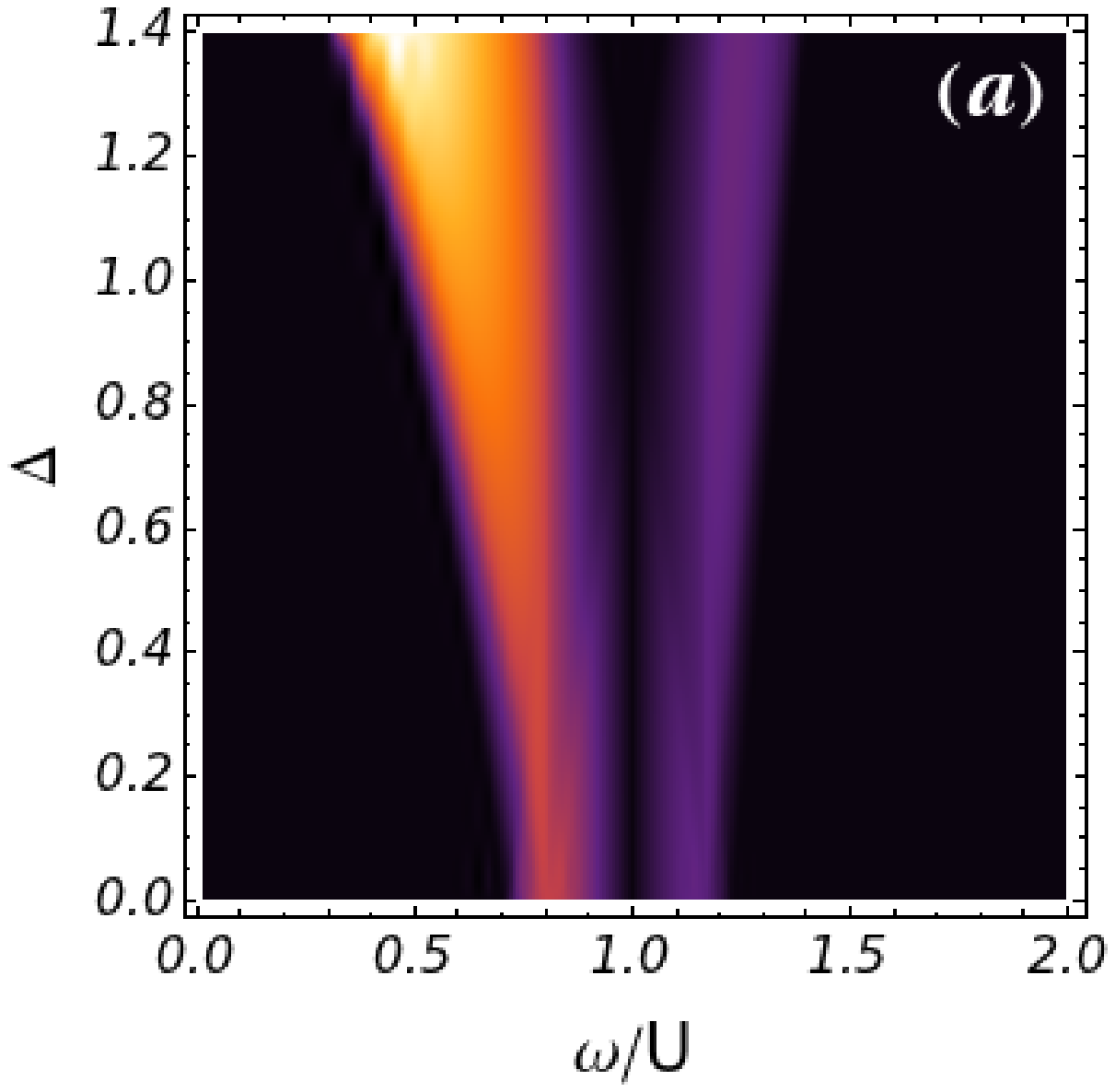}\includegraphics[scale=0.57]{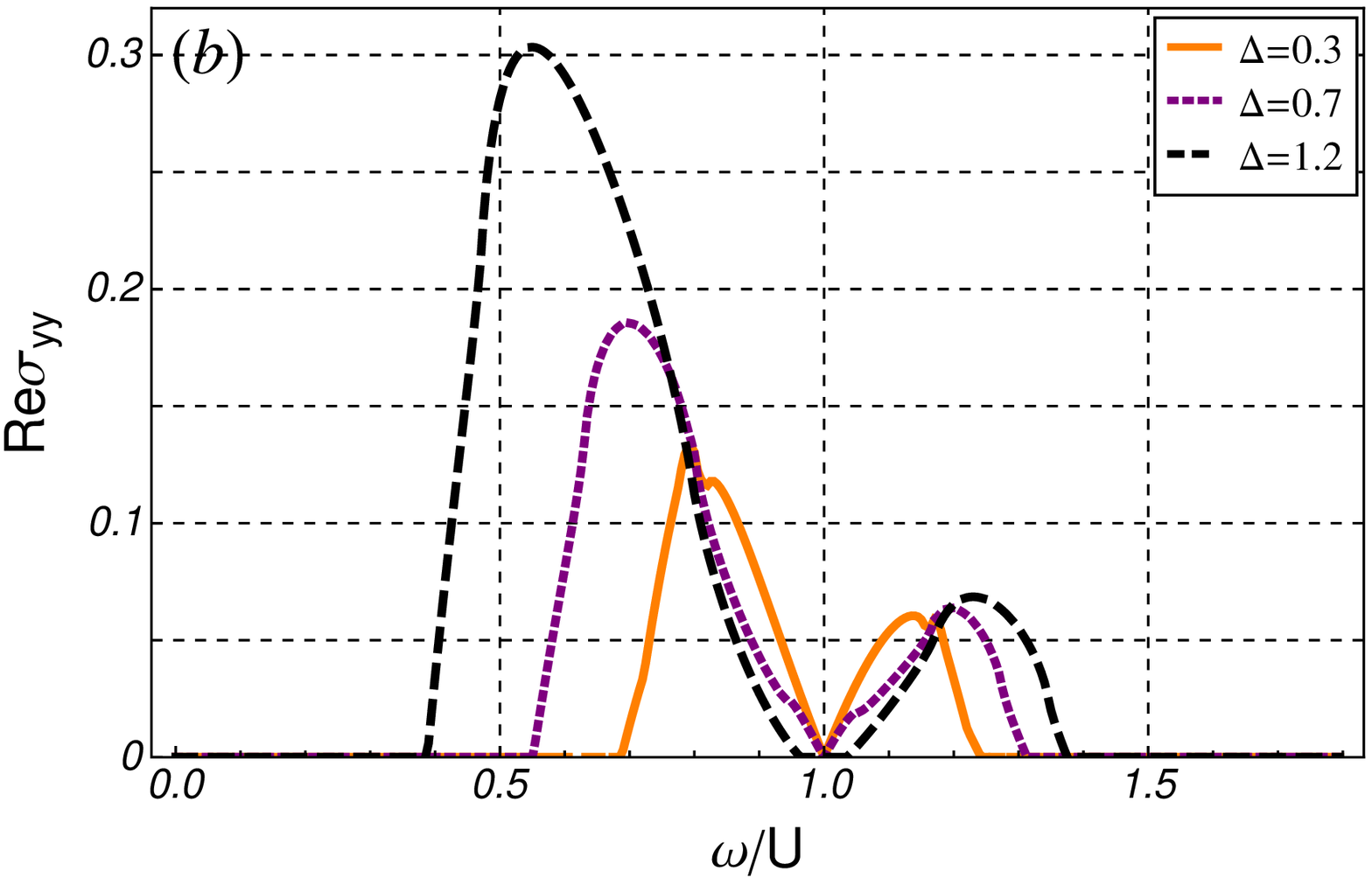}

\includegraphics[scale=0.55]{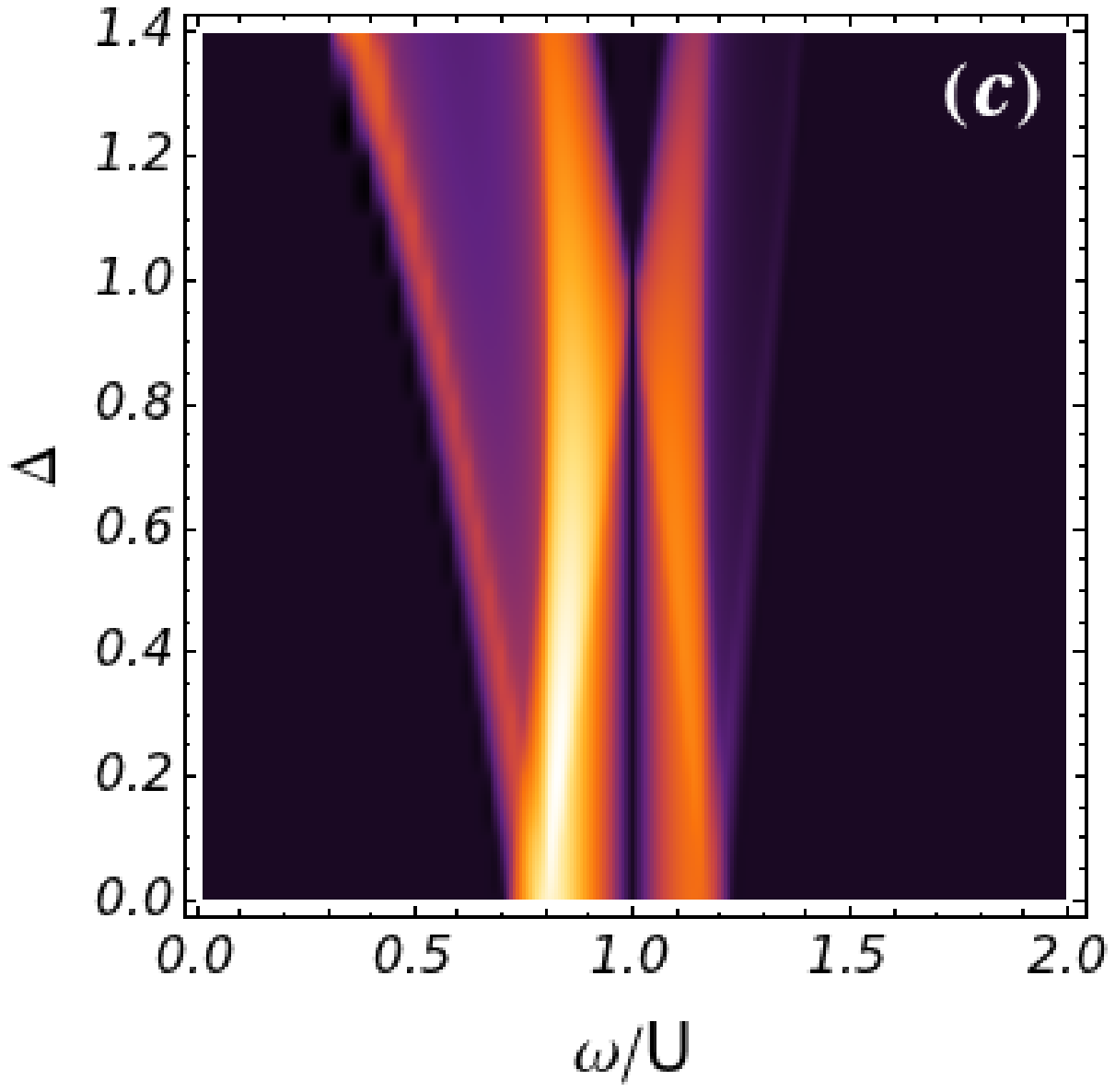}\includegraphics[scale=0.57]{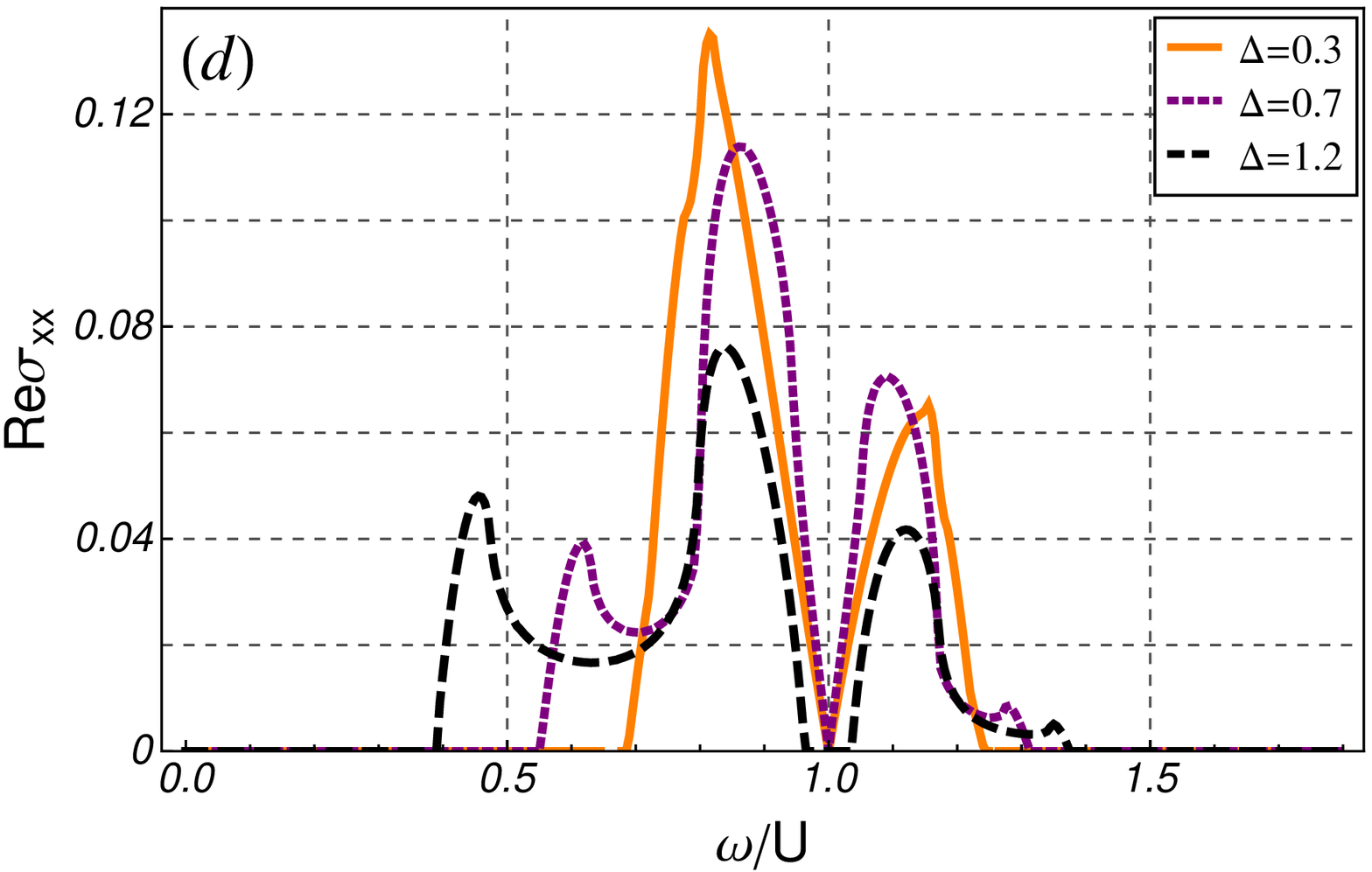}\caption{Analogous physical situation as in Fig. (\ref{fig:f-0_with_Delta})
but with non-zero amplitude of uniform magnetic field (half magnetic-flux
per unit cell).\label{fig:Frist-(second)-line}}
\end{figure*}

Fig. \ref{fig:f-0_with_Delta} and \ref{fig:Frist-(second)-line}
show the frequency dependence of OC for $f=0$ and $f=1/2$, respectively.
We are interested in $\sigma_{xx}^{\mathbf{A}_{0}}(\omega)$ and $\sigma_{yy}^{\mathbf{A}_{0}}(\omega)$
component in which the potential from site to site is varied in the
$x$ direction (to see the expressions used in the calculations of
OC see Appendix \ref{sub:appendix-staggered}). 

\textcolor{black}{The data presented in }Fig. \ref{fig:f-0_with_Delta}
and \ref{fig:Frist-(second)-line} imply a similar behavior of OC
when $\Delta$ parameter is alternated. For example, on the basis
of the staggered potential values with respect to its value for $\Delta=0$,
we conclude that it has greater impact on the $xx$ component of OC
than on $yy$ one. Besides the qualitative difference, we also observe
a smaller amplitude of the OC in direction $x$ than in that perpendicular
to $x$ within $xy$ plane. This behavior could be simply attributed
to the variation in the potential in this particular direction (i.e.
$x$). While the $yy$ component of the OC does not exhibit any special
difference along $y$ axis (for chosen strip of sites the $\Delta$
is constant). Moreover, we show that the increase in $\Delta$ causes
broadening\textcolor{black}{{} a frequency dependence of }OC. 

In agreement with the conclusion drawn in Sec. \ref{sub:Landau-gauge}
the insulator behavior of the OC for $\omega=U$ (Fig. \ref{fig:Frist-(second)-line})
is still maintained for $f=1/2$. In contrast to the situation with
no magnetic field (see Fig.\ref{fig:f-0_with_Delta}) the gap naturally
arises when the $\Delta$ parameter exceeds one (the single-particle
spectrum also exhibits a similar behavior). This gap-like behavior
is indeed observed in the quasi-particle energy dispersion $E_{q}^{\alpha\pm}(\mathbf{k};p)$
presented in Fig. \ref{fig:quasi} $d$ and $e$. Interestingly, the
weight of the OC close to $\omega=U$ for $f=1/2$ is greater than
for $f=0$. 

It is worth noting that the complex behavior of the OC in the direction
of applied uniaxially staggered potential gains pronounced peaks.
This should be easily observed in the EAR experiment (see Sec. \ref{sub:Connection-to-experiment}).

\subsection{Connection to experiment\label{sub:Connection-to-experiment}}

Recently, A. Tokuno and T. Giamarchi \cite{Tokuno2011} have proposed
a spectroscopic technique for cold atoms which is able to extract
current-current correlation function $\Pi_{xx}(\omega)$ (Eq. (\ref{eq:paramagnetic})).
This function is proportional to OC $\sigma_{xx}^{\mathbf{A}_{0}}(\omega)$
(see Eq. (\ref{eq:general_form_of_conductivity})), which offers a
possibility to probe the transport phenomena in the thermodynamic
limit. Using the energy absorption rate techniques (EAR) such a goal
could be achieved by phase modulation of optical lattice.

\begin{figure*}[th]
\includegraphics[scale=0.34]{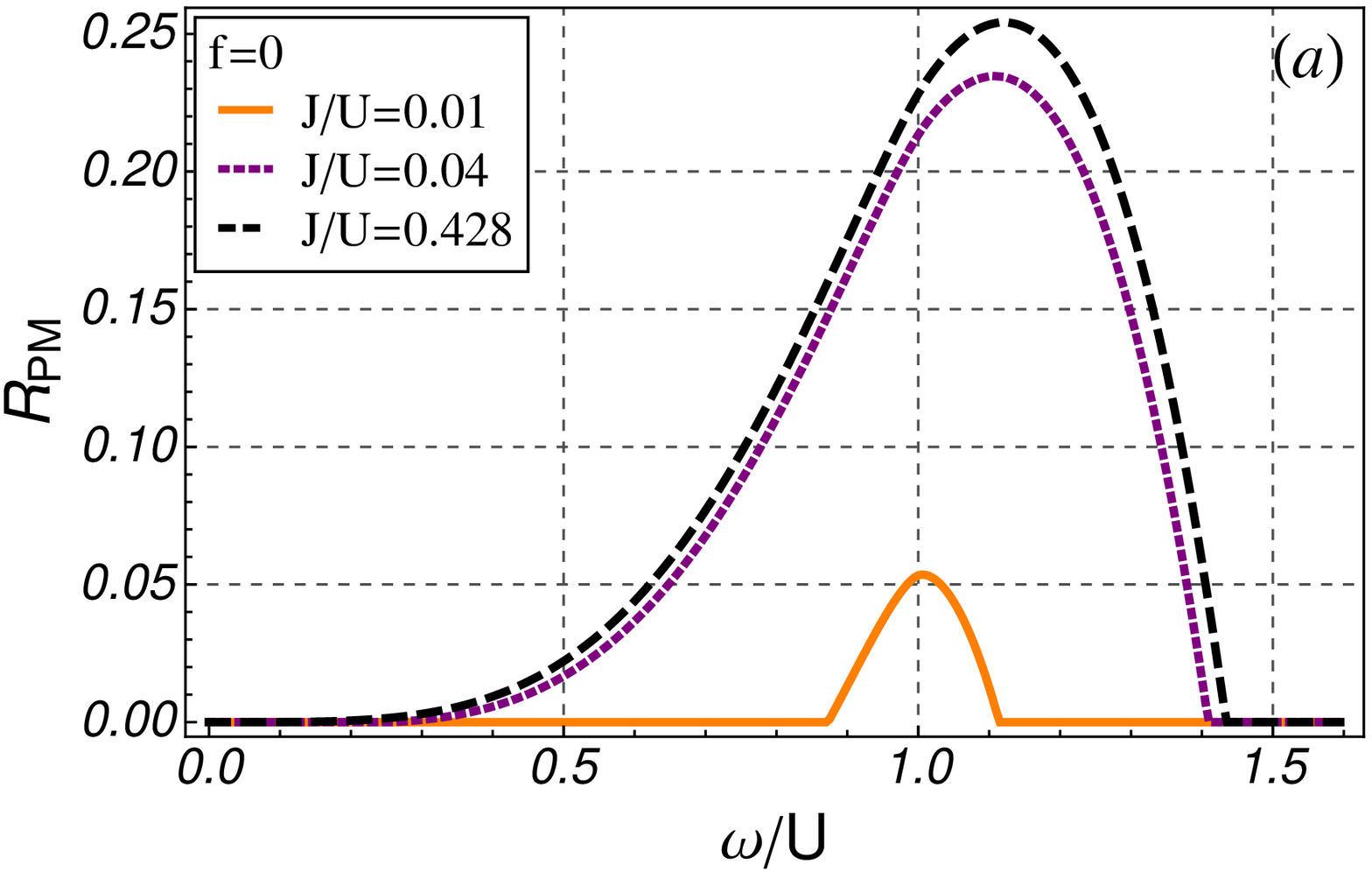}\includegraphics[scale=0.34]{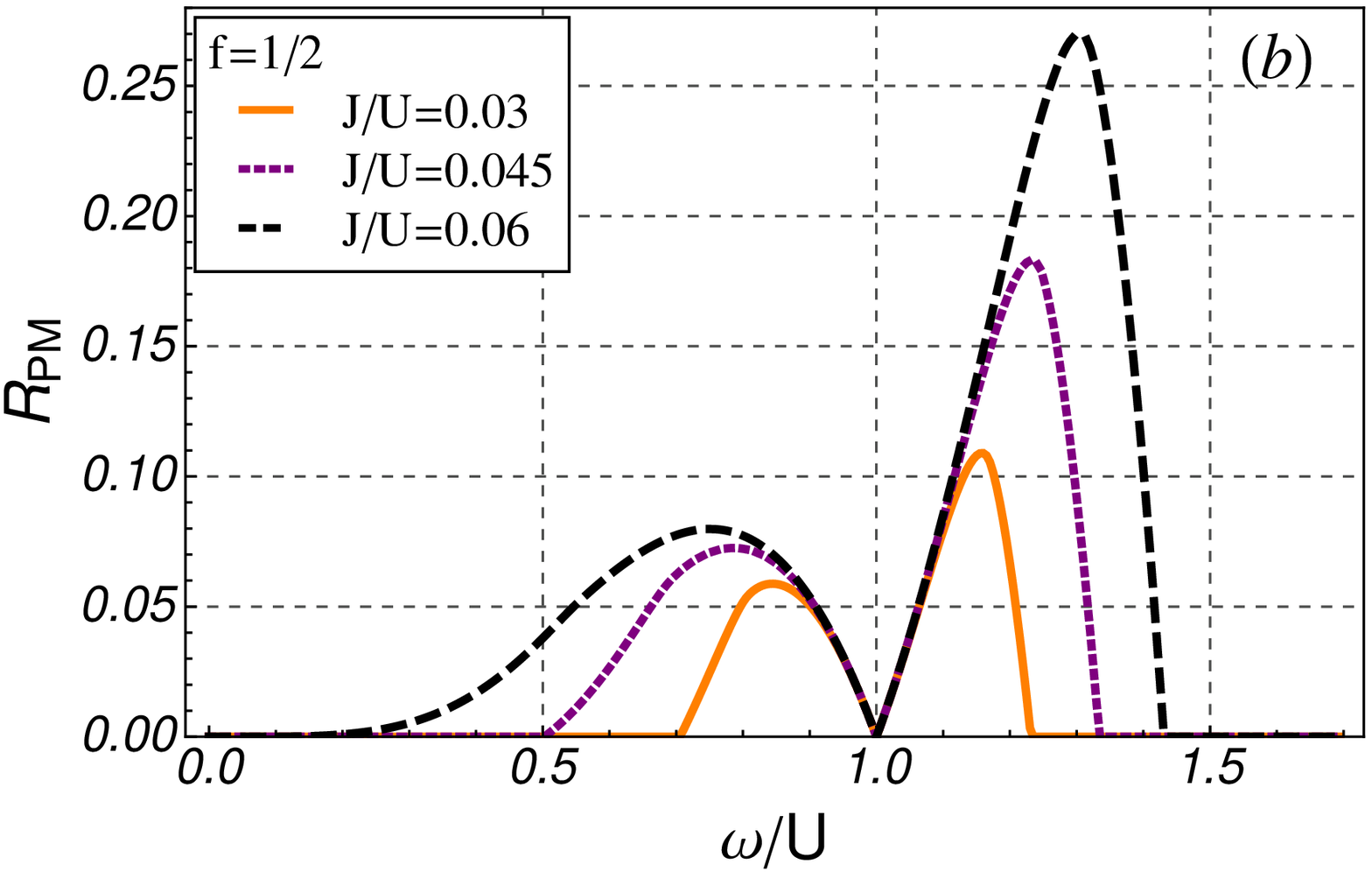}\includegraphics[scale=0.34]{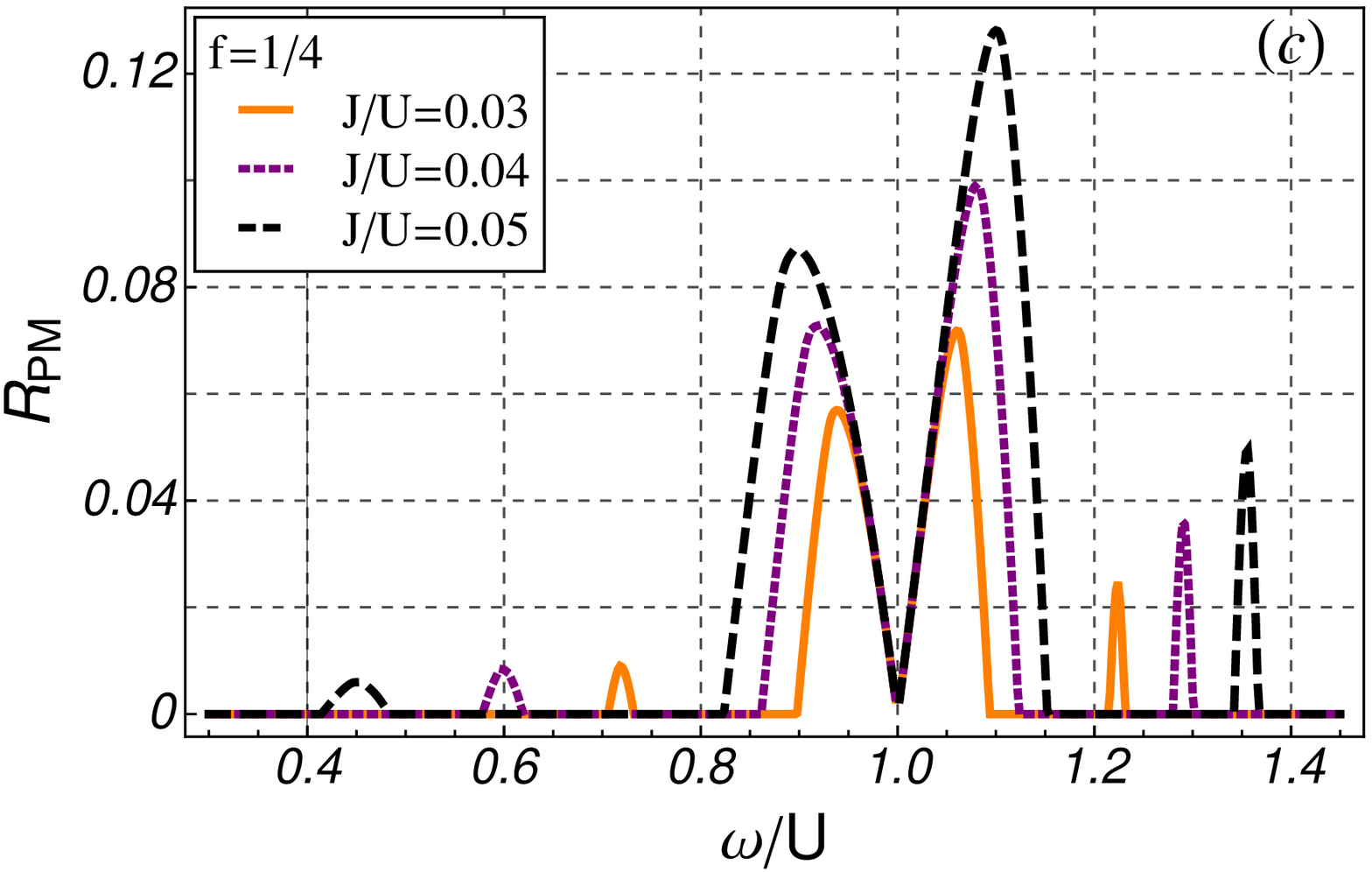}

\caption{Imaginary part of current-current correlation function in two-dimensional
square lattice in Mott phase (first lobe) is sketched in $\tilde{\sigma}_{Q}f_{x}^{2}$
units. From left - $f=0$ (a), $f=\frac{1}{2}$ (b), $f=\frac{1}{4}$
(c). Figures are plotted at zero temperature for different values
of $J/U$. \label{fig:EAR}}
\end{figure*}

Namely, a vector potential $\mathbf{A}_{0}$ could be created in different
experimental configurations \cite{Aidelsburger2011,Aidelsburger2013,Struck2011}.
In our work we investigate the Landau gauge $\mathbf{A}_{0}=B(0,x,0)$
which generates a uniform magnetic field. Such a uniform field has
been generated recently in \cite{Aidelsburger2013} but using another
type of effective vector potential. On the other hand, a small perturbing
vector potential $\mathbf{A}'$ could be generated by phase modulation
of optical lattice \cite{Tokuno2011}. If we consider a 2D system
with $x$ and $y$ axes, we can generate a synthetic electric field
$E\hat{x}$ by modulating the phase in the $x$ direction. This situation
could be mathematically inferred from the exchange of a stationary
optical lattice potential $V_{\textrm{op}}(\mathbf{r})=\cos^{2}\left(k_{x}x\right)+\cos^{2}\left(k_{y}y\right)$
to a time dependent one $V_{\textrm{op}}(\mathbf{r},t)$ where $x\rightarrow x-f_{x}\cos\left(\omega t\right)$
($f_{x}$ is the strength of modulation which should be much smaller
than a lattice constant). Such a phase modulation (PM) could be realized
by e.g. a recently proposed phase controller \cite{Sadgrove2011}.
This leads us to the expression where EAR is given by \cite{Tokuno2011}
\begin{equation}
R_{\textrm{PM}}(\omega)=-\frac{1}{2}\omega^{3}f_{x}^{2}\Im\tilde{\Pi}_{xx}(\omega)\;,
\end{equation}
in which $\tilde{\Pi}_{xx}(\omega)$ corresponds to $\Pi_{xx}(\omega)$
from Eq. (\ref{eq:paramagnetic}) but with exchange $e^{*}\rightarrow M$
($M$ is effective mass of an atom) \cite{Tokuno2011}. To ensure
a linear response regime and no dynamical phase transition the condition
$\omega f_{x}\ll1$ should be satisfied \cite{Eckardt2005,Drese1997}.
The calculation of $R_{\textrm{PM}}(\omega)$ is made following a
procedure similar to that used in the OC case where the current-current
correlation function $\Pi_{xx}(i\omega)$ was also considered (see
Eq. (\ref{eq:general_form_of_conductivity})). Fig. (\ref{fig:EAR})
presents a plot of $R_{\textrm{PM}}/\tilde{\sigma}_{Q}f_{x}^{2}$
for the uniform magnetic field of a strength $f=0,\;1/2,\;1/4$ on
the square lattice in the zero temperature limit (here $\tilde{\sigma}_{q}=M^{2}/h$
is an effective quantum resistance). We see that the factor $\omega^{3}$
changes significantly the weight of the response in comparison to
the OC given in Fig. \ref{fig:Optical-conductivity-in} and therefore
the higher frequency peaks give a grater contribution to the absorbed
energy rate. The similarity of the shape of current-current correlation
function and tight-binding dispersion in a strong magnetic field could
be an indirect method of checking the Hofstadter spectrum in the BHM
system \cite{Powell2010,Hofstadter1976a} where the center of the
band is located around $\omega=U$. 

The plots of EAR for the OC with staggered potential will be analogous
to the case of a uniform field (discussed above) so we omit its graphical
representation.

Summarizing, the EAR technique is directly related to OC and may act
as a probe of transport phenomena using ultra-cold quantum gases.
It is worth pointing out that the phase modulation is independent
of the strength of the lattice potential in contrast to the amplitude
modulation method \cite{Tokuno2011}.

\subsection{Critical conductivity at the MI-SF phase boundary\label{sub:Critical-conductivity}}

\subsubsection{The uniform field}

Up to now the problem of critical conductivity in 2D on the Mott insulator
- superfluid phase boundary in strong magnetic field has been rarely
studied because of its complexity. The amplitude of the hopping term
with a complex factor (see Eq. (\ref{eq:actionprzed3})) is the reason
why in Ref. \cite{Cha1994} instead of BHM at integer filling, the
frustrated XY model was investigated using the MC numerical method.
Another approach to the problem of critical conductivity in magnetic
field has been proposed in Ref. \cite{Nishiyama2001}, but the magnetic
field considered there was weaker (i.e. $f=1/20$) than that the field
we studied and their calculations were performed in the hard-core
limit (the author used exact diagonalization method to overcome the
difficulties related to the complex hopping term). Within the approach
presented in this paper we can perform analytical analysis and study
critical behavior of conductivity in a much wider range of magnetic
fields. Namely, we show that for commensurate value of $f$, critical
conductivity depends only on the number of minima located in the first
reduced magnetic Brillouin zone.

To describe the critical conductivity at the tip of the Mott lobe
\cite{Sinha2011} we compute the OC (Eq. (\ref{eq:cond_A final form}))
close to the phase boundary. To do that, in the following calculations,
we consider only the real part of $\sigma_{xx}^{\mathbf{A}_{0}}(\omega)$
that gives a finite frequency contribution, namely the part of OC
which consists of the current-current correlation function 
\begin{eqnarray}
 &  & \textrm{Re}\tilde{\sigma}_{xx}^{\mathbf{A}_{0}}(i\omega)=-\textrm{Re}\left\{ \frac{2\pi}{R_{q}}\frac{1}{\hbar\beta N}\frac{1}{\hbar\omega}\frac{1}{\hbar}\sum_{\mathbf{k}n}\sum_{\alpha=0}^{q-1}\times\right.\label{eq:sigma_pb}\\
 &  & \left.\left[\partial_{k_{x}}\epsilon_{q}^{\alpha}(\mathbf{k};p)\right]^{2}\mathcal{G}_{\alpha\alpha}^{d}(\mathbf{k},i\omega_{n})\mathcal{G}_{\alpha\alpha}^{d}(\mathbf{k},i\omega_{n}+i\omega)\right\} \;,\nonumber 
\end{eqnarray}
where $R_{q}=h/\left(e^{*}\right)^{2}=\sigma_{Q}^{-1}$ is quantum
resistance (for Cooper pair $e^{*}=2e$ and $R_{q}\approx6,45\; k\Omega$)
and we restore the constant $\hbar$ to introduce it in quantum resistance
$R_{q}$. Eq. (\ref{eq:sigma_pb}) also contains the singular part
of OC, but since we are interested in the Mott phase, we neglect this
contribution further on.

Now, using the effective action from Eq. (\ref{eq:action-eff}), and
applying the Ginzburg-Landau (GL) like method for calculation of critical
conductivity \cite{Kampf1993}, we evaluate Eq. (\ref{eq:sigma_pb})
in order to obtain its dependence on $f=p/q$ parameter. Following
this procedure, we expand the action Eq. (\ref{eq:action-eff}) to
the second order in frequency using the expression
\begin{equation}
1-\epsilon_{q}^{\alpha}(\mathbf{k};p)G_{0}(i\omega_{n})=a_{\mathbf{k}}-b_{\mathbf{k}}i\hbar\omega_{n}-c_{\mathbf{k}}\left(i\hbar\omega_{n}\right)^{2}\;,
\end{equation}
with
\begin{equation}
a_{\mathbf{k}}=1-\epsilon_{q}^{\alpha}(\mathbf{k};p)G_{0}(i\omega_{n}=0),\label{eq:ak}
\end{equation}
\[
b_{\mathbf{k}}=\epsilon_{q}^{\alpha}(\mathbf{k};p)\partial_{r}G_{0}(r)|_{r=0}\,,\quad c_{\mathbf{k}}=\frac{1}{2}\epsilon_{q}^{\alpha}(\mathbf{k};p)\partial_{r}^{2}G_{0}(r)|_{r=0}\;.
\]
Next, in calculating the critical conductivity within the GL action,
we should assume the proper ground state behavior. From all set of
$q-$band energy dispersion, we choose the lowest one which correctly
reproduces the phase transition. This band contains $q$ GL modes
in the first magnetic Brillouin zone (MBZ) \cite{Sinha2011}, which
allows description of the critical behavior of the BHM close to the
phase boundary. Going further we perform the summation over Matsubara
frequencies and take the limit $T\rightarrow0$, which reduces Eq.
(\ref{eq:sigma_pb}) to 
\begin{equation}
\textrm{Re}\tilde{\sigma}_{xx}^{\mathbf{A}_{0}}(\omega)_{\textrm{}}=\frac{\pi^{2}}{R_{q}}\frac{1}{N}\sum_{\mathbf{k}\mathbf{Q}}\frac{\left(\mathbf{k}-\mathbf{Q}\right)^{2}}{m_{eff}^{2}\Delta_{\mathbf{k}}^{2}}\frac{2J^{2}}{(\epsilon_{q}^{\alpha}(\mathbf{k};p))^{2}}\delta\left(\omega^{2}-\frac{\Delta_{\mathbf{k}}^{2}}{c_{\mathbf{k}}^{2}}\right),
\end{equation}
where $\Delta_{\mathbf{k}}=\sqrt{b_{\mathbf{k}}^{2}+4a_{\mathbf{k}}c_{\mathbf{k}}}$,
$m_{eff}=1/\left.\partial_{k_{x}}^{2}\epsilon_{q}^{\alpha}(\mathbf{k};p)\right|{}_{\mathbf{k}=\mathbf{Q}}$
and $\mathbf{Q}$ are locations of the minima in MBZ. Close to the
phase boundary only the momenta around $\mathbf{Q}$ bring a contribution
to conductivity, therefore if the minimum of $\epsilon_{q}^{\alpha}(\mathbf{k};p)$
is located at $\mathbf{k}=\mathbf{Q}$ we can simply expand $a_{\mathbf{k}}$
from Eq. (\ref{eq:ak}) to the second order 
\begin{equation}
a_{\mathbf{k}}\approx\frac{\left(\mathbf{k}-\mathbf{Q}\right)^{2}}{2zJm_{eff}}
\end{equation}
where $z=\epsilon_{q}^{\alpha}(\mathbf{Q};p)$ for chosen $p,\; q$.
In further calculations we assume that $\partial_{k_{x}}^{2}\epsilon_{q}^{\alpha}(\mathbf{k};p)=\partial_{k_{y}}^{2}\epsilon_{q}^{\alpha}(\mathbf{k};p)$.
Finally, for the point in the phase diagram which is close to the
tip of the lobe, we get
\begin{equation}
\textrm{Re}\tilde{\sigma}_{xx}^{\mathbf{A}_{0}}(\omega)_{\textrm{}}=\frac{\pi}{8R_{q}}\sum_{\mathbf{Q}}\frac{\omega^{2}-\left(\frac{b_{\mathbf{Q}}}{c_{\mathbf{\mathbf{Q}}}}\right)^{2}}{\omega^{2}}\Theta\left(\omega^{2}-\frac{b_{\mathbf{Q}}^{2}}{c_{\mathbf{\mathbf{Q}}}^{2}}\right)\;,\label{eq:conductivity final}
\end{equation}
where $\Theta(x)$ is a step function being non-zero for $x\geqslant0$.
The above expression describes the behavior of optical conductivity
close to the phase transition. It has non-vanishing amplitude when
the applied frequency is equal to $b_{\mathbf{Q}}/c_{\mathbf{\mathbf{Q}}}$
or is higher than this value. Denoting $\sigma^{*}=\pi\sigma_{Q}/8=\pi/8R_{q}$
and considering the tip of the lobe, where $b_{\mathbf{Q}}=0$ is
the critical conductivity, $\textrm{Re}\tilde{\sigma}_{xx}^{\mathbf{A}_{0}}(\omega)\equiv\sigma_{c,f}$
takes the simple form
\begin{equation}
\sigma_{c,f}=q\sigma^{*}\;.\label{eq:critical_cond}
\end{equation}
It is important to notice that theory presented here is valid for
the second order phase transition, e.g. this condition is satisfied
for the cases $f=1/2,\;1/3$ \cite{Cha1994}.

\begin{table}[t]
 \begin{center}   
\begin{tabular}{p{1cm} p{2cm}  p{1.9cm} p{1.5cm} p{1.6cm} }     \hline  \hline 
\multicolumn{1}{c}{f} &\multicolumn{4}{c}{normalized critical conductivity $\sigma_{c,f}\textrm{/}\sigma_{c,0}$ } \\ \cline{2-5}
 & MKF (here) & XY model  &  MC & JJA \newline experiment\\ \hline     
1/2 & 2 & 2 & 1,82 & $\approx 2$ \\     
1/3 & 3 &  & 2,91 & $\approx 3$ \\
1/q & q &  &  & $\approx q$ \\    \hline \hline   
\end{tabular} 
\end{center}

\caption{Comparison of critical conductivity for different magnetic field configuration
calculations \cite{Granato1990,Grason2006,Cha1991,Cha1994} and experimental
measurements \cite{Zant1992,VanderZantHS1996}.\label{tab:Critical-conductivity-for}}

\end{table}

To discuss the result from Eq. (\ref{eq:critical_cond}) we firstly
recall that for the simplest case $q=1$ where there is no magnetic
field, the result for $q=1$ confirms the results presented in Refs.
\cite{Cha1991,Kampf1993,VanOtterloA1993}. With $p/q=1/2$ we have
$\sigma_{c,1/2}=2\sigma^{*}$. This result agrees with the analytical
solution given in Refs. \cite{Grason2006,Granato1990}, where the
XY model was used. Also in the MC study for $f=1/2$ \cite{Cha1994}
a value of critical conductivity is $1,82$ times higher than for
$f=0$ \cite{Cha1991}. Results of an experiment conducted in Josephson
junction arrays \cite{Zant1992,VanderZantHS1996} also show a similar
behavior. If we consider $f=1/3$, it qualitatively agrees with the
Monte Carlo result in Ref. \cite{Cha1994}, where Cha and Grivin obtained
$2,91$ higher value of critical conductivity than in the absence
of a magnetic field. The authors of the experimental work, in Refs.
\cite{Zant1992,VanderZantHS1996} also discuss such a scenario. It
is worth adding that the authors of Ref. \cite{Granato1990} have
speculated about a similar result for $\sigma_{c,1/3}$ i.e. $\sigma_{c,1/3}=3\sigma_{c,0}$.
Namely, they have suggested that at least for low order rationals
of $p/q$ critical conductivity could satisfy Eq. (\ref{eq:critical_cond})
but they have carried out explicit calculation only for $f=1/2$.
In contrast, we showed this behavior by analytical methods for arbitrary
$f$ within a clear mathematical framework. The above considerations
are summarized in Table \ref{tab:Critical-conductivity-for}. It is
worth mentioning that in three dimensions a trivial solution $\sigma_{c,f}=0$
is obtained, known before only for the case of $f=0$ (e.g, see \cite{Kampf1993}).

For Josephson junction arrays (JJA) it seems that the critical conductivity
is proportional to $q/p$ \cite{VanderZantHS1996} but this linear
behavior is inferred from a small number of experimental points with
a large error margin. Therefore, such a dependence is still an open
question. Moreover, in (JJA) we should take into account that arrays
are not perfect and their parameters differ through the network. Also,
the measurements are performed at finite temperatures. For example,
if we consider a disorder we should expect that this effect suppresses
the value of critical conductivity \cite{Sorensen1992,Cha1994}. It
is worth mentioning here that also long range interactions which we
neglected could have a significant impact \cite{Sorensen1992}. Besides,
it seems that $\sigma_{c,f}$ should depend on $p$. Hence our results
(\ref{eq:critical_cond}) within the approximations used in this paper
should be at least appropriate for $p\approx1$.

\subsubsection{The uniform field (f=1/2) with uniaxially staggered potential}

To show the importance of translation symmetry breaking by the uniaxially
staggered potential in a special case $f=1/2$ \cite{Delplace2010},
we analyze the critical behavior of conductivity at the phase boundary.

\begin{figure}[t]
\includegraphics[scale=0.33]{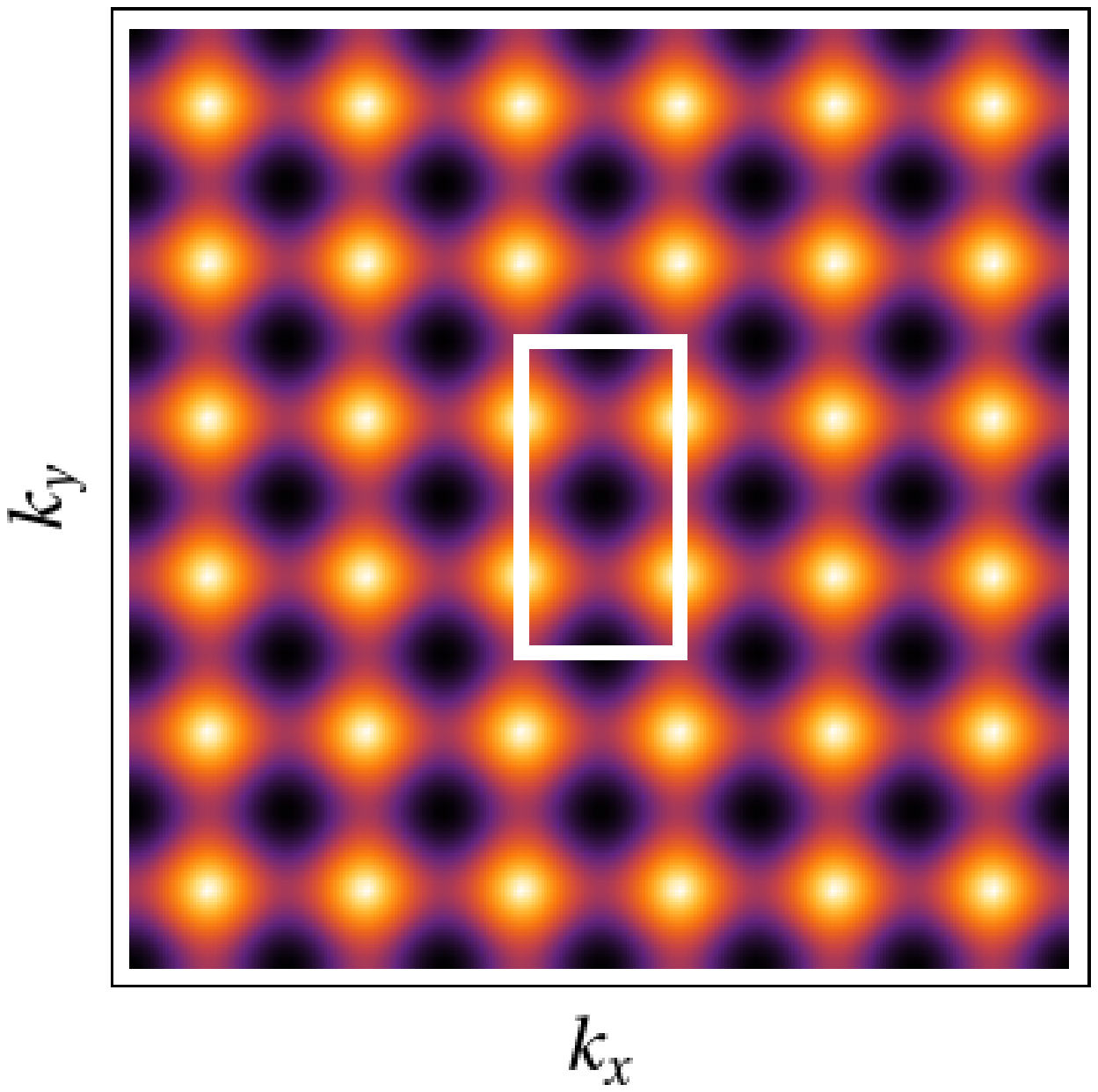}\includegraphics[scale=0.33]{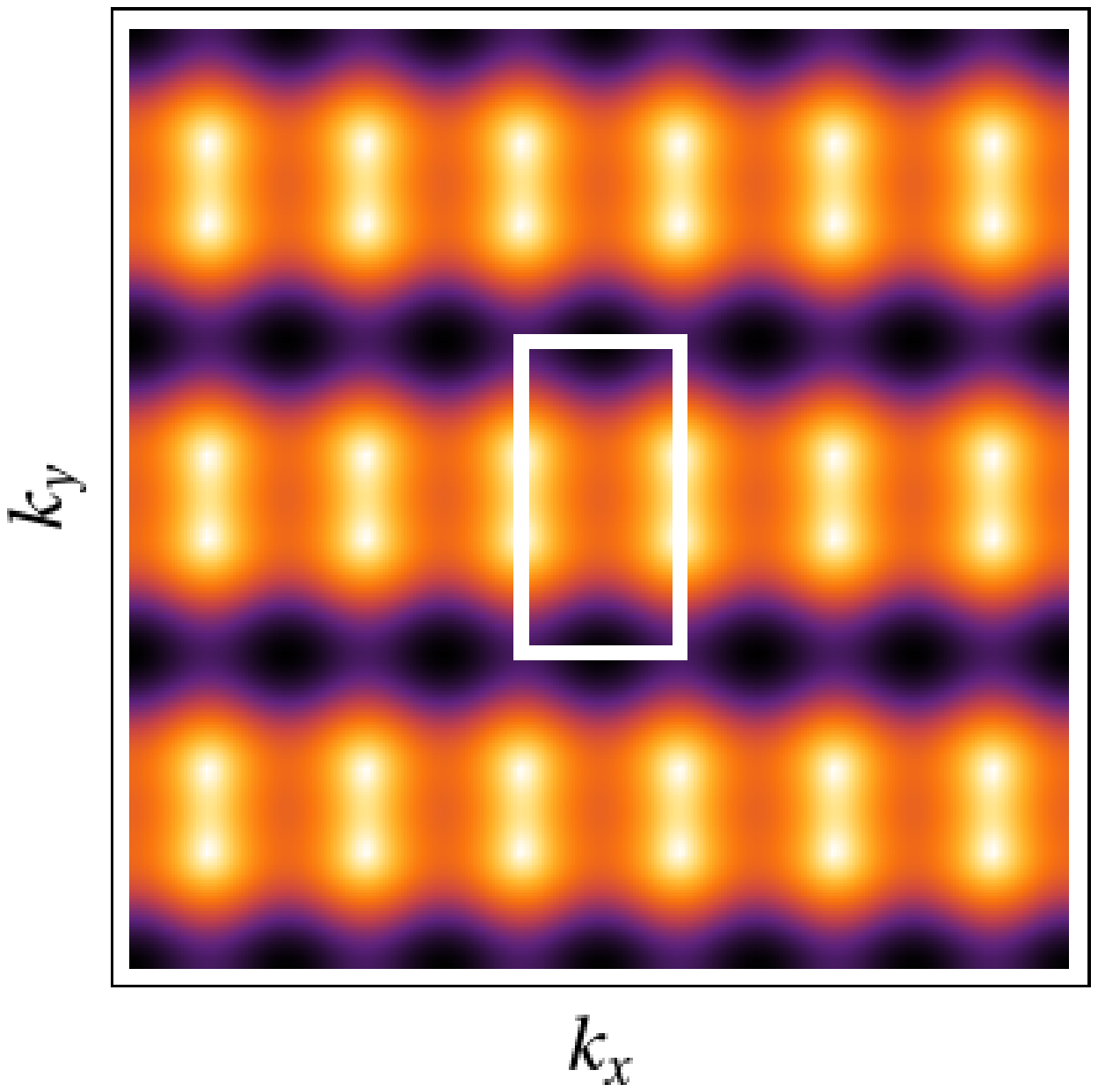}

\caption{Above we sketch density plot of the lowest band energy in the Mott
phase (white rectangle describe first magnetic Brillouin zone (MBZ)).
The range from darker to brighter color is assigned to the lowest
and highest value of energy spectrum respectively. Pictures have $\Delta=0$
and $\Delta=0.7$ (from left). We see explicitly disappearance one
of the two minima (black color) on the center of the MBZ.\label{fig:density-plot}}

\end{figure}

Following the same procedure for the critical conductivity which led
us to Eq. (\ref{eq:critical_cond}) we simply observe that for $\Delta=0$
(for definition of parameter $\Delta$ see Sec. \ref{sub:Staggered-potential})
we get $\sigma_{c,f}=2\sigma^{*}$. But for $\Delta>0$, the situation
is changed significantly. Analysis of the spectrum of quasi-particles
in the Mott phase reveals that one of the two minima disappears in
the first magnetic Brillouin zone for the non-zero value of $\Delta$
(see Fig. \ref{fig:density-plot}). Therefore, there exists only the
one lowest energy Ginzburg-Landau mode which effectively recovers
the critical conductivity as when there is no magnetic field. Hence,
such an abrupt change in the critical value of $\sigma_{c,f}$ from
$2\sigma^{*}$ to $\sigma^{*}$when the staggered potential is turned
on, could be an interesting effects on its own right\textcolor{black}{.
In addition, no variation in the critical conductivity for $f=0$
was observed.}

The best options to study the presented effects in experiments that
could be made nowadays is the use of optical lattices with ultra-cold
atoms whose high controllability provides a better road to connect
experiment and theory. The continuous progress in ultra-cold quantum
gases in the near future also will allow getting into the temperature
regime where high precision measurement of critical conductivity will
be possible. This can verify the above results and allow omission
of additional effects which occur in standard solid state devices
like Josephson junction arrays. The possibility of such measurements
has been very recently discussed in Ref. \cite{Chen2013}.

The analysis made in Sec. \ref{sub:Critical-conductivity} permits
a better understanding of the superconductor-insulator phase transition
mechanism. To illustrate this, we shortly explain below the fact that
the critical conductivity should be affected by the applied magnetic
field. Namely, we know that the scenario of critical resistance (i.e.
$1/\sigma_{c,f}$) is assigned to the vortex and boson flowing through
the system at the critical transition point \cite{Cha1991}. Such
a description is possible from the duality transformation \cite{Fisher1990,Fisher1990a}
(e.g. for superconductor-insulator transition induced in a fermionic
system these bosons are Cooper pairs with short coherence length \cite{Cha1991}).
Therefore, application of a magnetic field at least affects the behavior
of vortices, which changes the critical resistance. In our calculations
the magnetic field is effectively incorporated into the theory through
the tight binding dispersion relation $\epsilon_{q}^{\alpha}(\mathbf{k};p)$
which gives $q-$bands spectrum with $q$ minima in the lowest energy
level, what finally changes $\sigma_{c,f}$. If we additionally consider
the staggered potential, an analogous prediction could be made.

\section{Summary}

The analysis of conductivity in BHM in a strong magnetic field is
challenging problem due to complex hopping term. In particular, up
to now its optical dependence was out of reach in Monte Carlo study.
Therefore our theory expands the area in which numerical methods are
used.

Namely, we have proposed the magnetic Kubo formula which is valid
for an arbitrary flux pattern where commensurability effects of a
magnetic field are included. Within this framework, we have calculated
the optical conductivity in the Mott phase of the Bose Hubbard model
and considered its critical value. To check our results we have proposed
to compare them with presently available experiment in ultra-cold
quantum gases in which current-current correlation function in a uniform
magnetic field could be probed. Such a connection of experiments and
theory could open a new avenue to study transport phenomena in highly
controllable magnetic field where geometry of the lattice can be easily
manipulated. Moreover, for the case of critical conductivity we have
shown its dependence on the topology of the single-particle spectrum
and obtained solution which are in good agreement with presently available
numerical and experimental data.

The method presented here can be extended over many-body systems in
which the strong magnetic field plays a significant role.

\begin{acknowledgments}
The work was supported by (Polish) National Science Center Grant No.
DEC-2011/01/D/ST2/02019 (A.S.S., T.P.P.).
\end{acknowledgments}

\section{Appendix\label{sec:densityofstates}}

Here we present the quasi-particle energy spectrum and densities of
states for conductivity (DOSc) in 2D described in Sec. \ref{sub:Optical-conductivity-1-2-4}.
We use $\mathcal{K}$ and $\mathcal{E}$ to denote the first and second
complete elliptic integral, respectively.

\subsection{Uniform magnetic field for a two-dimensional square lattice\label{sub:appendix-uniform}}

\begin{widetext}

\subsubsection{f=0}

The dispersion relation for the square lattice is
\begin{equation}
\epsilon_{1}^{\alpha}(\mathbf{k};0)=-2J\left(\cos k_{x}+\cos k_{y}\right),
\end{equation}
where the DOSc (Eq. (\ref{eq:DOS-cond}) is given by \cite{Belkhir1994}

\begin{equation}
\rho_{1}^{0}(E;0)=\frac{4\Theta\left(4-|E|\right)}{\pi^{2}}\left[\mathcal{E}\left(\sqrt{1-\left(\frac{E}{4}\right)^{2}}\right)-\left(\frac{E}{4}\right)^{2}\mathcal{K}\left(\sqrt{1-\left(\frac{E}{4}\right)^{2}}\right)\right].
\end{equation}

\subsubsection{f=1/2}

The dispersion relation for $f=1/2$ is built from two sub-bands \cite{Hasegawa1989}
$\pm2\sqrt{\cos^{2}k_{x}+\cos k_{y}^{2}}$ and consequently the DOSc
is

\begin{equation}
\rho_{2}^{\alpha}(E;1)=\frac{4\Theta_{\alpha}}{\pi^{2}|E|}\left[\mathcal{E}\left(\sqrt{1-\left(\frac{E^{2}-4}{4}\right)^{2}}\right)-\left(\frac{E^{2}-4}{4}\right)^{2}\mathcal{K}\left(\sqrt{1-\left(\frac{E^{2}-4}{4}\right)^{2}}\right)\right]\;,
\end{equation}
where $\Theta_{\alpha}$ is non-zero step function within each q-band.

\subsubsection{f=1/4}

The form of dispersion relation $f=1/4$ is expressed by four sub-bands
\cite{Hasegawa1989} $\pm\sqrt{4-\sqrt{12+2\cos\left(4k_{x}\right)+2\cos\left(4k_{y}\right)}}$
and $\pm\sqrt{4+\sqrt{12+2\cos\left(4k_{x}\right)+2\cos\left(4k_{y}\right)}}$,
then we get

\begin{equation}
\rho_{4}^{\alpha}(E;1)=\frac{4\Theta_{\alpha}}{\pi^{2}\left|E^{2}-4\right||E|}\left[\mathcal{E}\left(\sqrt{1-\left(\frac{4-8E^{2}+E^{4}}{4}\right)^{2}}\right)-\left(\frac{4-8E^{2}+E^{4}}{4}\right)^{2}\mathcal{K}\left(\sqrt{1-\left(\frac{4-8E^{2}+E^{4}}{4}\right)^{2}}\right)\right]\;.
\end{equation}

\end{widetext}

\subsection{Uniform magnetic field for a two-dimensional square lattice with
uniaxially staggered potential\label{sub:appendix-staggered}}

\subsubsection{f=0}

The form of tight binding dispersion has two sub-bands \cite{Delplace2010}
$-2J\cos k_{y}\pm2J\sqrt{\cos^{2}k_{x}+\Delta^{2}}$ and the appropriate
DOSc is given by

- for $xx$ component of optical conductivity $\sigma_{xx}^{\mathbf{A}_{0}}(\omega)$:
\begin{equation}
\rho_{2}^{\alpha}(E;1)=\frac{2\Theta_{\alpha}}{\pi^{2}}\int_{0}^{1}dx\,\frac{x^{2}}{x^{2}+\Delta^{2}}\frac{\sqrt{1-x^{2}}}{\sqrt{1-\left(\frac{E}{2}\pm\sqrt{x^{2}+\Delta^{2}}\right)^{2}}}\;,
\end{equation}

- for $yy$ component of optical conductivity $\sigma_{yy}^{\mathbf{A}_{0}}(\omega)$:

\begin{equation}
\rho_{2}^{\alpha}(E;1)=\frac{2\Theta_{\alpha}}{\pi^{2}}\int_{0}^{1}dx\,\frac{\sqrt{1-\left(\frac{E}{2}\pm\sqrt{x^{2}+\Delta^{2}}\right)^{2}}}{\sqrt{1-x^{2}}}\;.
\end{equation}

\subsubsection{f=1/2}

The form of tight binding dispersion has two sub-bands \cite{Delplace2010}
$\pm2\sqrt{\cos^{2}k_{x}+\left(\cos k_{y}-\Delta\right)^{2}}$ and
the appropriate DOSc is given by

- for $xx$ component of optical conductivity $\sigma_{xx}^{\mathbf{A}_{0}}(\omega)$:

\begin{equation}
\rho_{2}^{\alpha}(E;1)=\frac{4\Theta_{\alpha}}{\pi^{2}|E|}\int_{-1-\Delta}^{1-\Delta}dx\,\frac{\sqrt{\left(\frac{E}{2}\right)^{2}-x^{2}}\sqrt{1-\left(\frac{E}{2}\right)^{2}+x^{2}}}{\sqrt{1-\left(x+\Delta\right)^{2}}}\;,
\end{equation}

- for $yy$ component of optical conductivity $\sigma_{yy}^{\mathbf{A}_{0}}(\omega)$:
\begin{equation}
\rho_{2}^{\alpha}(E;1)=\frac{4\Theta_{\alpha}}{\pi^{2}|E|}\int_{-1-\Delta}^{1-\Delta}dx\,\frac{x^{2}}{\sqrt{\left(\frac{E}{2}\right)^{2}-x^{2}}}\frac{\sqrt{1-\left(x+\Delta\right)^{2}}}{\sqrt{1-\left(\frac{E}{2}\right)^{2}+x^{2}}}\;.
\end{equation}

\bibliographystyle{apsrev}
\bibliography{library}

\end{document}